\documentclass[aps,prd,twocolumn,showpacs,superscriptaddress,groupedaddress,nofootinbib]{revtex4-2}  % for review and submission
\usepackage{cancel}
\usepackage{graphicx} % Required for inserting images
\usepackage[T1]{fontenc}
\usepackage{lmodern} 
\usepackage{graphicx}  % needed for figures
\usepackage{dcolumn}   % needed for some tables
\usepackage{bm}        % for math
\usepackage{amssymb}   % for math
\usepackage{amsmath}
\usepackage{dsfont}
\usepackage{bbold}
\usepackage{array}
\usepackage{float}
\usepackage{makecell}
\usepackage{braket}
\usepackage{longtable}
\usepackage{supertabular,booktabs}
\usepackage{soul}

\usepackage{titlesec} %for hyperref to detect names of sections etc
\usepackage[colorlinks,citecolor=blue,urlcolor=blue,hypertexnames=true]{hyperref}
\usepackage{subfigure}
\usepackage[usenames,dvipsnames,svgnames,table]{xcolor} % use color 

\renewcommand{\arraystretch}{2}

\newcommand{\be}{\begin{equation}}
\newcommand{\ee}{\end{equation}}
\newcommand{\bea}{\begin{eqnarray}}
\newcommand{\eea}{\end{eqnarray}}
\newcommand{\bml}{\begin{subequations}}
\newcommand{\eml}{\end{subequations}}
\newcommand{\bfig}{\begin{figure}}
\newcommand{\efig}{\end{figure}}

\newcommand{\slashed}{\hspace{-1.1ex}/}

\newcommand{\bmat}{\begin{pmatrix}}
\newcommand{\emat}{\end{pmatrix}}
\usepackage{graphicx, slashed,booktabs, color, multirow, float,
amsfonts, bbold, mathtools, sidecap, tikz, bm,enumitem}
\usepackage{multirow}
\usepackage{bbding}
\usepackage{titlesec}
\usepackage{hyperref}
\usepackage{wasysym}
\usepackage{amssymb}% http://ctan.org/pkg/amssymb
\usepackage{pifont}% http://ctan.org/pkg/pifont

\renewcommand{\leq}{\leqslant}
\renewcommand{\geq}{\geqslant}

\usepackage[dvipsnames, usenames]{xcolor}

\definecolor{linkcolor}{rgb}{0.55, 0.13, .32}

\definecolor{oucrimsonred}{rgb}{0.6, 0.0, 0.0}
\definecolor{persianblue}{rgb}{0.11, 0.22, 0.73}
\definecolor{forestgreen}{rgb}{0.13,0.35,0.13}
\definecolor{lightgray}{rgb}{0.83, 0.83, 0.83}
 \hypersetup{colorlinks, citecolor=oucrimsonred, linkcolor=persianblue, urlcolor=oucrimsonred}
\definecolor{cornellred}{rgb}{0.7, 0.11, 0.11}
\definecolor{navyblue}{rgb}{0.0, 0.0, 0.5}
\definecolor{amethyst}{rgb}{0.6, 0.4, 0.8}
\definecolor{yellow}{rgb}{1.0, 1.0, 0.0}
\definecolor{firebrick}{rgb}{0.7, 0.13, 0.13}
\definecolor{tangerineyellow}{rgb}{1.0, 0.8, 0.0}
\definecolor{deepfuchsia}{rgb}{0.76, 0.33, 0.76}
\definecolor{amber}{rgb}{1.0, 0.75, 0.0}
\definecolor{VioletRed4}{rgb}{0.55, 0.13, .32}
\definecolor{indiagreen}{rgb}{0.07, 0.53, 0.03}
\definecolor{VioletRed4}{rgb}{0.55, 0.13, .32}

\usepackage{hyperref}
\usepackage{graphics, appendix, afterpage, makecell} 
\usepackage{bbold}
\usepackage{tikz}
\usepackage{adjustbox}

\usepackage{tcolorbox}

\definecolor{oucrimsonred}{rgb}{0.6, 0.0, 0.0}
\definecolor{persianblue}{rgb}{0.11, 0.22, 0.73}
\definecolor{forestgreen}{rgb}{0.13,0.35,0.13}
\definecolor{lightgray}{rgb}{0.83, 0.83, 0.83}
 \hypersetup{colorlinks, citecolor=oucrimsonred, linkcolor=persianblue, urlcolor=oucrimsonred}
\definecolor{cornellred}{rgb}{0.7, 0.11, 0.11}
\definecolor{navyblue}{rgb}{0.0, 0.0, 0.5}
\definecolor{amethyst}{rgb}{0.6, 0.4, 0.8}
\definecolor{yellow}{rgb}{1.0, 1.0, 0.0}
\definecolor{firebrick}{rgb}{0.7, 0.13, 0.13}
\definecolor{tangerineyellow}{rgb}{1.0, 0.8, 0.0}
\definecolor{deepfuchsia}{rgb}{0.76, 0.33, 0.76}
\definecolor{amber}{rgb}{1.0, 0.75, 0.0}
\definecolor{VioletRed4}{rgb}{0.55, 0.13, .32}
\definecolor{indiagreen}{rgb}{0.07, 0.53, 0.03}
\definecolor{VioletRed4}{rgb}{0.55, 0.13, .32}
\newcommand{\nn}{\nonumber}

\definecolor{oucrimsonred}{rgb}{0.6, 0.0, 0.0}
\newcommand\vertarrowbox[3][6ex]{%
  \begin{array}[t]{@{}c@{}} #2 \\
  \left\uparrow\vcenter{\hrule height #1}\right.\kern-\nulldelimiterspace\\
  \makebox[0pt]{\scriptsize#3}
  \end{array}%
}

\definecolor{mtcolor}{rgb}{.8,.3,.1}

\definecolor{violachiaro}{rgb}{1,0.6,1}

\definecolor{gbcolor}{rgb}{.43,.22,.12}
 
\definecolor{gbcolor2}{rgb}{.9,.2,.6}
\definecolor{gbcolor3}{rgb}{.3,.2,.6}

\definecolor{verdechiaro}{rgb}{0.6,1,0.6}
\definecolor{giallochiaro}{rgb}{1,1,0.6}
\definecolor{bluscuro}{rgb}{0.15, 0.2, 0.9}
\definecolor{verdes}{rgb}{0.1, 0.5, 0.1}%
\definecolor{tangerineyellow}{rgb}{1.0, 0.8, 0.0}
\definecolor{smokyblack}{rgb}{0.06, 0.05, 0.03}

\definecolor{americanrose}{rgb}{1.0, 0.01, 0.24}
\definecolor{cobalt}{rgb}{0.0, 0.28, 0.67}
\definecolor{brandeisblue}{rgb}{0.0, 0.44, 1.0}
\definecolor{mycolor}{rgb}{0.0, 0.0, 0.5}%navyblue
\definecolor{oxfordblue}{rgb}{0.0, 0.13, 0.28}
\definecolor{azure}{rgb}{0.0, 0.5, 1.0}
\definecolor{turquoiseblue}{rgb}{0.0, 1.0, 0.94}
\newtcolorbox{mynewbox}[1]{colback=white!5!white,colframe=azure!75!black,fonttitle=\bfseries,title=#1}
\newtcolorbox{mybox}{colback=mycolor!5!white,colframe=azure!75!black}
\newtcolorbox{mynamedbox}[1]{colback=mycolor!5!white,colframe=azure!75!black,title=#1}
\definecolor{venetianred}{rgb}{0.78, 0.03, 0.08}
\newtcolorbox{mynamedbox1}[1]{colback=venetianred!5!white,colframe=venetianred!80!black,title=#1}
\newtcolorbox{mynamedbox2}[1]{colback=azure!5!white,colframe=azure!80!black,title=#1}

\definecolor{rossocorsa}{rgb}{0.83, 0.0, 0.0}

\tikzset{->-/.style={decoration={
  markings,
  mark=at position #1 with {\arrow{>}}},postaction={decorate}}}
\tikzset{-<-/.style={decoration={
  markings,
  mark=at position #1 with {\arrow{<}}},postaction={decorate}}} 
\def\be{\begin{equation}}
\def\ee{\end{equation}}
\def\ba{\begin{eqnarray}}
\def\ea{\end{eqnarray}}

\def\L*{{\cal L}_*}
\def\L{\mathcal{L}}
\def\({\left(}
\def\){\right)}

\def\nn{\nonumber}

\def\<{\langle}
\def\>{\rangle}

\def\cs2{c_{s}^{2}}

 \def\be   {\begin{equation}}   \def\ee   {\end{equation}}
 \def\ba   {\begin{array}}      \def\ea   {\end{array}}
 \def\bea  {\begin{eqnarray}}   \def\eea  {\end{eqnarray}}
 \def\bean {\begin{eqnarray*}}  \def\eean {\end{eqnarray*}}
\titleclass{\subsubsubsection}{straight}[\subsection]

\newcounter{subsubsubsection}[subsubsection]
\renewcommand\thesubsubsubsection{\thesubsubsection.\arabic{subsubsubsection}}
\titleformat{\subsubsubsection}
  {\normalfont\normalsize\bfseries}{\thesubsubsubsection}{1em}{}
\titlespacing*{\subsubsubsection}
{0pt}{3.25ex plus 1ex minus .2ex}{1.5ex plus .2ex}

\makeatletter
\renewcommand\paragraph{\@startsection{paragraph}{5}{\z@}%
  {3.25ex \@plus1ex \@minus.2ex}%
  {-1em}%
  {\normalfont\normalsize\bfseries}}
\renewcommand\subparagraph{\@startsection{subparagraph}{6}{\parindent}%
  {3.25ex \@plus1ex \@minus .2ex}%
  {-1em}%
  {\normalfont\normalsize\bfseries}}
\def\toclevel@subsubsubsection{4}
\def\toclevel@paragraph{5}
\def\toclevel@paragraph{6}
\def\l@subsubsubsection{\@dottedtocline{4}{7em}{4em}}
\def\l@paragraph{\@dottedtocline{5}{10em}{5em}}
\def\l@subparagraph{\@dottedtocline{6}{14em}{6em}}
\makeatother

\usepackage{titlesec} %for hyperref to detect names of sections etc
\usepackage[colorlinks,citecolor=blue,urlcolor=blue,hypertexnames=true]{hyperref}
\usepackage{subfigure}
\usepackage[usenames,dvipsnames,svgnames,table]{xcolor}  
\usepackage{tikzsymbols}
\usepackage{natbib}
\usepackage{float}

\usepackage{tikz,xcolor,hyperref}

\usepackage{slashed}

\definecolor{lime}{HTML}{A6CE39}
\DeclareRobustCommand{\orcidicon}{
	\begin{tikzpicture}
	\draw[lime, fill=lime] (0,0) 
	circle [radius=0.2] 
	node[white] {{\fontfamily{qag}\selectfont \tiny ID}};
	\draw[white, fill=white] (-0.0625,0.095) 
	circle [radius=0.007];
	\end{tikzpicture}
	\hspace{-2mm}
}

\foreach \x in {A, ..., Z}{\expandafter\xdef\csname orcid\x\endcsname{\noexpand\href{https://orcid.org/\csname orcidauthor\x\endcsname}
			{\noexpand\orcidicon}}
}
\usepackage{bbold}
\usepackage{tikz}
\usepackage{adjustbox}
\usepackage{tcolorbox}
\usepackage{enumitem}
\usepackage{amsfonts}

\setlist[itemize,1]{label=$\times$}
\setlist[itemize,2]{label=$\checkmark$}
\setlist[itemize,3]{label=$\diamond$}
\setlist[itemize,4]{label=$\bullet$}

\begin{document}
\title{\Large \textcolor{Sepia}{
Obviating PBH overproduction for SIGWs generated by Pulsar Timing Arrays in loop corrected EFT of bounce
%Evading PBH overproduction for SIGWs generated by Pulsar Timing Arrays with non-Gaussianity in single-field Galileon inflation
}}
\author{\large Sayantan Choudhury\orcidA{}${}^{1}$}
\email{sayantan\_ccsp@sgtuniversity.org,  \\ sayanphysicsisi@gmail.com (Corresponding author)} 
 \author{\large Siddhant Ganguly\orcidB{}{}${}^{2}$}
\email{ms22100@iisermohali.ac.in }
\author{\quad\quad\quad\quad\quad\quad\quad\quad\quad\quad\quad\quad\quad\quad\quad\quad\quad\quad\quad\quad\quad\quad\quad\quad\quad\quad\quad\quad\quad\quad\quad\quad\quad\quad\quad\quad\quad\quad\quad\quad\large Sudhakar~Panda\orcidD{}${}^{1,3}$}
\email{panda@niser.ac.in}
\author{\large Soumitra SenGupta${}^{4}$}
\email{tpssg@iacs.res.in, soumitraiacs@gmail.com}
\author{\large Pranjal Tiwari\orcidC{}${}^{2}$}
\email{ms22104@iisermohali.ac.in}

\affiliation{ ${}^{1}$Centre For Cosmology and Science Popularization (CCSP),\\
        SGT University, Gurugram, Delhi- NCR, Haryana- 122505, India,}
 \affiliation{${}^{2}$Indian Institute of Science Education and Research, Mohali-140306, India}
\affiliation{${}^{3}$School of Physical Sciences,  National Institute of Science Education and Research, Bhubaneswar, Odisha - 752050, India.}
\affiliation{${}^{4}$School of Physical Sciences, Indian Association for the Cultivation of Science,
2A \& 2B Raja S.C Mullick Road, Kolkata-700032, India.}

\begin{abstract}
%%%%%%%%%%%%%%%%%%%%%%%%%%%%%%%%%%%%%%%%%%%

In order to unravel the present situation of the PBH overproduction problem, our study emphasizes the critical role played by the equation of state (EoS) parameter $w$ within the framework of effective field theory (EFT) of non-singular bounce. Our analysis focuses on a wide range of EoS parameter values that are still optimal for explaining the latest data from the pulsar timing array (PTA). As a result of our study, the most advantageous window, $0.31 \leq w \leq 1/3$, is identified as the location of a substantial PBH abundance, $f_{\rm PBH} \in (10^{-3},1)$ with large mass PBHs, $M_{\rm PBH}\sim {\cal O}(10^{-7}-10^{-3})M_{\odot}$, in the SIGW interpretation of the PTA signal. When confronted with PTA, we find that the overproduction avoiding circumstances are between $1\sigma-2\sigma$, while the EoS parameter lies inside the narrow window, $0.31<w\leq 1/3$. We propose a regularized-renormalized-resummed (RRR) scalar power spectrum that is large enough to produce EoS dependent scalar generated gravitational waves compatible with PTA evidence, while satisfying the perturbativity, causality, and unitarity criteria, within the range of $0.88 \leq c_{s} \leq 1$.

%%%%%%%%%%%%%%%%%%%%%%%%%%%%%%%%%%%%%%%%%%%
\end{abstract}

\maketitle
\section{Introduction}
Primordial black holes (PBHs) have garnered significant interest lately due to their association with generated gravitational waves and their prospective role as dark matter candidates. See refs. \cite{Zeldovich:1967lct,Hawking:1974rv,Carr:1974nx,Carr:1975qj,Chapline:1975ojl,Carr:1993aq,Choudhury:2011jt,Yokoyama:1998pt,Kawasaki:1998vx,Rubin:2001yw,Khlopov:2002yi,Khlopov:2004sc,Saito:2008em,Khlopov:2008qy,Carr:2009jm,Choudhury:2011jt,Lyth:2011kj,Drees:2011yz,Drees:2011hb,Ezquiaga:2017fvi,Kannike:2017bxn,Hertzberg:2017dkh,Pi:2017gih,Gao:2018pvq,Dalianis:2018frf,Cicoli:2018asa,Ozsoy:2018flq,Byrnes:2018txb,Ballesteros:2018wlw,Belotsky:2018wph,Martin:2019nuw,Ezquiaga:2019ftu,Motohashi:2019rhu,Fu:2019ttf,Ashoorioon:2019xqc,Auclair:2020csm,Vennin:2020kng,Nanopoulos:2020nnh,Inomata:2021uqj,Stamou:2021qdk,Ng:2021hll,Wang:2021kbh,Kawai:2021edk,Solbi:2021rse,Ballesteros:2021fsp,Rigopoulos:2021nhv,Animali:2022otk,Frolovsky:2022ewg,Escriva:2022duf,Ozsoy:2023ryl,Ivanov:1994pa,Afshordi:2003zb,Frampton:2010sw,Carr:2016drx,Kawasaki:2016pql,Inomata:2017okj,Espinosa:2017sgp,Ballesteros:2017fsr,Sasaki:2018dmp,Ballesteros:2019hus,Dalianis:2019asr,Cheong:2019vzl,Green:2020jor,Carr:2020xqk,Ballesteros:2020qam,Carr:2020gox,Ozsoy:2020kat,Baumann:2007zm,Saito:2008jc,Saito:2009jt,Choudhury:2013woa,Sasaki:2016jop,Raidal:2017mfl,Papanikolaou:2020qtd,Ali-Haimoud:2017rtz,Di:2017ndc,Raidal:2018bbj,Cheng:2018yyr,Vaskonen:2019jpv,Drees:2019xpp,Hall:2020daa,Ballesteros:2020qam,Carr:2020gox,Ozsoy:2020kat,Ashoorioon:2020hln,Papanikolaou:2020qtd,Wu:2021zta,Kimura:2021sqz,Solbi:2021wbo,Teimoori:2021pte,Cicoli:2022sih,Ashoorioon:2022raz,Papanikolaou:2022chm,Papanikolaou:2023crz,Wang:2022nml,ZhengRuiFeng:2021zoz,Cohen:2022clv,Cicoli:2022sih,Brown:2017osf,Palma:2020ejf,Geller:2022nkr,Braglia:2022phb,Frolovsky:2023xid,Aldabergenov:2023yrk,Aoki:2022bvj,Frolovsky:2022qpg,Aldabergenov:2022rfc,Ishikawa:2021xya,Gundhi:2020kzm,Aldabergenov:2020bpt,Cai:2018dig,Cheng:2021lif,Balaji:2022rsy,Qin:2023lgo,Riotto:2023hoz,Riotto:2023gpm,Papanikolaou:2022did,Choudhury:2011jt,Choudhury:2023vuj, Choudhury:2023jlt, Choudhury:2023rks,Choudhury:2023hvf,Choudhury:2023kdb,Choudhury:2023hfm,Bhattacharya:2023ysp,Choudhury:2023fwk,Choudhury:2023fjs,Choudhury:2024one,Choudhury:2024ybk,Choudhury:2024jlz,Choudhury:2024dei,Choudhury:2024aji,Harada:2013epa,Harada:2017fjm,Kokubu:2018fxy,Gu:2023mmd,Saburov:2023buy,Stamou:2023vxu,Libanore:2023ovr,Friedlander:2023qmc,Chen:2023lou,Cai:2023uhc,Karam:2023haj,Iacconi:2023slv,Gehrman:2023esa,Padilla:2023lbv,Xie:2023cwi,Meng:2022low,Qiu:2022klm,Mu:2022dku,Fu:2022ypp,Davies:2023hhn,Firouzjahi:2023ahg,Firouzjahi:2023aum, Iacconi:2023ggt,Davies:2023hhn,Jackson:2023obv,Riotto:2024ayo,Maity:2023qzw,Ragavendra:2024yfp,Papanikolaou:2024rlq,Papanikolaou:2024kjb,Banerjee:2021lqu,Choudhury:2023kam,Heydari:2021gea,Heydari:2021qsr,Heydari:2023xts,Heydari:2023rmq,Caravano:2024tlp,Banerjee:2022xft,Papanikolaou:2021uhe,Papanikolaou:2022hkg} for more details.
Several experiments such as NANOGrav15 \cite{NANOGrav:2023gor, NANOGrav:2023hde, NANOGrav:2023ctt, NANOGrav:2023hvm, NANOGrav:2023hfp, NANOGrav:2023tcn, NANOGrav:2023pdq, NANOGrav:2023icp, Inomata:2023zup}, EPTA \cite{EPTA:2023fyk, EPTA:2023sfo, EPTA:2023akd, EPTA:2023gyr, EPTA:2023xxk, EPTA:2023xiy, Lozanov:2023rcd}, PPTA \cite{Reardon:2023gzh, Reardon:2023zen, Zic:2023gta}, and CPTA \cite{Xu:2023wog} have found a Stochastic Gravitational Wave Background (SGWB) signal pervading the entire universe and reaching Earth from all directions. There have been several proposed explanations for this phenomenon, including the mergers of supermassive black holes, first-order phase transitions, cosmic strings, domain walls, inflation, and Scalar Induced Gravitational Waves (SIGW). See refs. \cite{Choudhury:2023fwk,Choudhury:2023fjs,Choudhury:2024one,Choudhury:2023hfm,Bhattacharya:2023ysp,Franciolini:2023pbf,Inomata:2023zup,Wang:2023ost,Balaji:2023ehk,HosseiniMansoori:2023mqh,Gorji:2023sil,DeLuca:2023tun,Choudhury:2023kam,Yi:2023mbm,Cai:2023dls,Cai:2023uhc,Huang:2023chx,Vagnozzi:2023lwo,Frosina:2023nxu,Zhu:2023faa,Jiang:2023gfe,Cheung:2023ihl,Oikonomou:2023qfz,Liu:2023pau,Liu:2023ymk,Wang:2023len,Zu:2023olm, Abe:2023yrw, Gouttenoire:2023bqy,Salvio:2023ynn, Xue:2021gyq, Nakai:2020oit, Athron:2023mer,Ben-Dayan:2023lwd, Madge:2023cak,Kitajima:2023cek, Babichev:2023pbf, Zhang:2023nrs, Zeng:2023jut, Ferreira:2022zzo, An:2023idh, Li:2023tdx,Blanco-Pillado:2021ygr,Buchmuller:2021mbb,Ellis:2020ena,Buchmuller:2020lbh,Blasi:2020mfx, Madge:2023cak, Liu:2023pau, Yi:2023npi,Vagnozzi:2020gtf,Benetti:2021uea,Inomata:2023drn,Lozanov:2023rcd,Basilakos:2023jvp,Basilakos:2023xof,Li:2023xtl,Domenech:2021ztg,Yuan:2021qgz,Chen:2019xse,Cang:2023ysz,Cang:2022jyc,Konoplya:2023fmh,Huang:2023chx,Ellis:2023oxs,Yu:2023jrs,Nassiri-Rad:2023asg,Heydari:2023rmq,Li:2023xtl,Chang:2023aba,Bernardo:2023jhs,Choi:2023tun,Elizalde:2023rds,Chen:2023bms,Nojiri:2023mbo,Domenech:2023jve,Liu:2023hpw,Huang:2023zvs,Oikonomou:2023bli,Cyr:2023pgw,Fu:2023aab,Kawai:2023nqs,Kawasaki:2023rfx,Maji:2023fhv,Bhaumik:2023wmw,He:2023ado,An:2023jxf,Zhu:2023lbf,Das:2023nmm,Roshan:2024qnv,Chen:2024fir,Chowdhury:2023xvy,Battista:2021rlh,Battista:2022hmv,Battista:2022sci,Battista:2023znv,DeFalco:2023djo,DeFalco:2024ojf,Domenech:2024rks,Iovino:2024uxp,Andres-Carcasona:2024wqk} for more details. That being said, there are worries over the overproduction of PBHs \cite{Franciolini:2023pbf,Franciolini:2023wun,Inomata:2023zup,LISACosmologyWorkingGroup:2023njw, Inui:2023qsd,Chang:2023aba, Gorji:2023ziy,Li:2023xtl, Li:2023qua,Firouzjahi:2023xke,Gorji:2023sil,Ota:2022xni,Raatikainen:2023bzk,Choudhury:2023fwk,Choudhury:2023fjs,Choudhury:2024one} that result from the development of these objects from heightened curvature disturbances in the very early cosmos.

Current Cosmic Microwave Background (CMB) observations allow us to probe universe on large scales, reaching back to the radiation-matter decoupling. Before this period, the universe was opaque to electromagnetic radiation preventing exploration using traditional methods. However, recent advancements in Gravitational Wave astronomy has provided a novel method to study the very early universe, as gravitational waves do not interact with intervening matter and can travel across the universe unimpeded. In this work, we would focus on the Scalar Induced Gravitational Waves to explain SGWB and concerns related to the PBH production associated with the frequency regime of the PTA signal (${\cal O}(10^{-9} - 10^{-6})$ Hz.). SIGWs are tensor perturbations formed due to the curvature perturbations in early universe, set against an otherwise flat FLRW background. Significant enhancements in these perturbations lead to the formation of overdense and underdense regions. If the density of some of the overdense regions exceed a certain threshold, they undergo gravitational collapse to form PBHs. To achieve these enhancements, an ultra-slow roll phase is required in addition to the usual slow roll phase. One of the problems arising from PBH formation due to enhanced curvature perturbation in the early universe is that of overproduction. Several theories predict that the abundance of PBHs can exceed that of dark matter, which raises a concern, as PBHs are one of the candidates for dark matter \cite{Ivanov:1994pa,Afshordi:2003zb,Frampton:2010sw,Carr:2016drx,Kawasaki:2016pql,Inomata:2017okj,Espinosa:2017sgp,Ballesteros:2017fsr,Sasaki:2018dmp,Ballesteros:2019hus,Dalianis:2019asr,Cheong:2019vzl,Green:2020jor,Carr:2020xqk,Ballesteros:2020qam,Carr:2020gox,Ozsoy:2020kat}. This is known as PBH Overproduction problem. There have been several proposed solutions to the problems. The details of these approaches can be found in \cite{Franciolini:2023pbf,Franciolini:2023wun,Inomata:2023zup,LISACosmologyWorkingGroup:2023njw, Inui:2023qsd,Chang:2023aba, Gorji:2023ziy,Li:2023xtl, Li:2023qua,Firouzjahi:2023xke,Gorji:2023sil,Ota:2022xni,Raatikainen:2023bzk,Choudhury:2023fwk,Choudhury:2023fjs,Choudhury:2024one}. 

As of right now, the PTA collaboration's latest work has mostly concentrated on the formation of these gravitational waves, identifying their primary source as the radiation-dominated period with an equation of state of $w=1/3$. Even if there are valid reasons to take into account such a scenario, current research has demonstrated that studying the primordial world with a constant EoS $w$ and propagation speed $c_{s}$ is valuable \cite{Liu:2023pau,Liu:2023hpw,Balaji:2023ehk,Domenech:2021ztg,Domenech:2020ers,Domenech:2019quo}. Approaches to deal with situations that may have happened before to the start of BBN are promised by this endeavor. We will be using the EoS parameter ($w$) to resolve this problem by parameterizing it to study the results where $w$ takes several values, as the relative energy density of SIGWs $(\Omega_{\rm GW})$ and PBH abundance relative to dark matter abundance $(f_{\rm PBH})$ depend on the EoS parameter. We also parameterize the propagation speed $c_{s}$ because it directly affects the peak amplitude of the power spectrum which in turn affects the value of $f_{\rm PBH}$. This paper investigates the case in which a non-adiabatic perfect fluid with a constant $w$ and constant propagation speed $c_{s}$ parameters dominates the early Universe's background. This fluid lies within the window $0.88\leq c_s\leq 1$, preserving the perturbativity, unitarity, and causality criteria simultaneously. Here, we operate in the linear regime, which does not take super-Hubble scale effects of non-Gaussianity into consideration. In this study, we also use the Press-Schechter formalism to determine the PBH mass fraction. It is demonstrated that employing the generic EoS formalism to prevent PBH overproduction is effective only when its values are close to the epoch that the EoS describes within a narrow window of favorable conditions, $0.31\leq w\leq 1/3$, and when the overdensity and ensuing Gaussian statistics are assumed to be linear.

%The SGWB signals lie in the range of ${\cal O}(10^{-9} - 10^{-6})$ Hz. Incidentally, this frequency range coincides with the range where PBH forming curvature perturbations can cause observable distortions in signals from various GW events.

In this paper we build upon the work done in \cite{Choudhury:2023vuj,Choudhury:2023jlt,Choudhury:2023rks}, which uses large quantum corrections to the power spectrum and also use the regularization-renormalization-resummation (RRR) procedure to get rid of any divergences brought up by quantum-loop corrections. This procedure leads to a no-go theorem which prohibits the formation of PBHs above $M_{\rm PBH}\sim {\cal O}(10^2){\rm gm}$ in single field inflation. This implies that if one wishes to produce PBHs of solar mass range, one cannot achieve successful inflation as the number of e-folding halts around ${\cal O}(25)$. Several alternatives such as stochastic inflation \cite{Choudhury:2024jlz}, Multiple Sharp Transition induced inflation \cite{Bhattacharya:2023ysp,Choudhury:2023fjs}, Galelion Inflation \cite{Choudhury:2023hvf,Choudhury:2023kdb,Choudhury:2023hfm,Choudhury:2024one}, and the theory of non-singular bounce \cite{Choudhury:2024dei} have been proposed to evade the strong no-go theorem \cite{Choudhury:2023vuj,Choudhury:2023jlt,Choudhury:2023rks}. All these theories were based on the effective field theory (EFT) setup \cite{Weinberg:2008hq,Cheung:2007st,Choudhury:2017glj,Choudhury:2024ybk,Choudhury:2024jlz,Choudhury:2021brg,Adhikari:2022oxr,Banerjee:2021lqu,Naskar:2017ekm,Choudhury:2015pqa,Choudhury:2015eua,Choudhury:2015zlc,Choudhury:2015hvr,Choudhury:2014sua,Choudhury:2013iaa,Choudhury:2013jya,Choudhury:2013zna,Choudhury:2011sq,Choudhury:2011jt,Choudhury:2012yh,Choudhury:2012whm,Choudhury:2014sxa,Choudhury:2014uxa,Choudhury:2014kma,Choudhury:2016cso,Choudhury:2016pfr,Choudhury:2017cos,Bohra:2019wxu,Akhtar:2019qdn,Choudhury:2020yaa,Choudhury:2021tuu,Choudhury:2016wlj,Cabass:2022avo,Cai:2016thi,Cai:2017tku,Agarwal:2012mq,Piazza:2013coa,Delacretaz:2016nhw,Salcedo:2024smn,Colas:2023wxa,Senatore:2010wk,Noumi:2012vr,Tong:2017iat,Noumi:2012vr,Arkani-Hamed:2015bza,Kim:2021pbr,Baumann:2018muz,Choudhury:2018glz,Hongo:2018ant,Baumann:2017jvh,An:2017hlx,Gong:2017yih,Liu:2016aaf}. We use a theory of non-singular bounce followed by inflation including a USR phase in between which is preferably short-lived in order to control the enhancement of the scalar perturbation. Addition of contraction and bounce phases evades the strong no-go theorem. Implementing EFT of non-singular bouncing framework offers a novel and promising way to get over the limitations resulting from the aforementioned one-loop corrections and enable the generation of PBHs having mass, $M_{\rm PBH}\leq {\cal O}(M_{\odot})$, which is compatible with the microlensing experiments \cite{Niikura:2017zjd,Niikura:2019kqi,EROS-2:2006ryy} along with GW signal obtained from PTA data.

Specifically, we will concentrate on the PBH generation within the mass range that the reported NANOGrav15 signal predicts. Additionally, we will produce the spectrum of the GW energy density $\Omega_{\rm GW}h^{2}$, which when combined can stay compatible with the most current PTA data. We have now added the EoS as a new factor to our research in order to support our decision and look into potential further features. We will address PBH production in the context of EFT, where the USR regime provides the required enhancement in the curvature perturbations to enable PBH formation \cite{Kristiano:2022maq,Riotto:2023hoz,Choudhury:2023vuj,Choudhury:2023jlt,Choudhury:2023rks,Choudhury:2023hvf,Choudhury:2023kdb,Choudhury:2023hfm,Bhattacharya:2023ysp,Choudhury:2023fjs,Banerjee:2021lqu,Kristiano:2023scm,Riotto:2023gpm,Firouzjahi:2023ahg,Firouzjahi:2023aum,Firouzjahi:2023bkt,Franciolini:2023lgy,Cheng:2023ikq,Tasinato:2023ukp,Tasinato:2023ioq,Motohashi:2023syh,Jackson:2023obv,Davies:2023hhn,Iacconi:2023ggt,Mu:2023wdt,Domenech:2023dxx,Ahmadi:2023qcw,Dalianis:2023pur,Tada:2023rgp,Kawaguchi:2023mgk,Tomberg:2023kli,Ragavendra:2023ret,Zhai:2023azx}. This is achieved by combining an SR/USR/SR-like setup with a contracting and bouncing phase. The overproduction of PBH within the PTA-accessible frequency range has lately come to light as a serious problem that requires careful attention to resolve. In this study, we investigate the idea that the formation threshold for PBHs is an EoS-dependent quantity. Since we are unable to pinpoint the exact state of the universe prior to it becoming dominated by radiation, we investigate a range of EoS values where $w \ne 1/3$ allows us to constrain the mass fraction of the PBH to give us a significant abundance, preventing overproduction. Non-linear effects are still not a part of our methodology. Investigating the impact of include the equation of state in the non-linear domain using the gradient expansion technique is required for a thorough study to address the overproduction issue. Despite the difficulties this job presents, we want to get started on it soon. However, we expect that the outcomes we have found here won't be much altered.

The outline of this work is as follows: In section \ref{EFTBasic}, we develop the basic setup for the effective field theory (EFT) of bounce to build the second-order perturbations for the Goldstone modes. In section \ref{TreePS} we give the expression for the tree-level power spectrum constructed via mode quantization. Further in section \ref{RRR} we discuss the Regularized-Renormalized-Resummed (RRR) loop corrected Power Spectrum in detail which is the prime component for our further discussions. In section \ref{PBH}, we use the Press-Schechter formalism to derive the expressions for the mass fraction($M_{\rm PBH}$) and PBH abundance ($f_{\rm PBH}$) as a function of $w$. In section \ref{SIGW}, we build an expression for the relative energy density ($\Omega_{\rm GW}$) for the RRR power spectrum and plot it using numerical techniques along with the observed NANOGrav15 and EPTA data for different values of $w$ and $c_s$. In section \ref{OP}, we give a overview of the PBH Overproduction problem and how we intend to resolve it with the help of general EoS. In section \ref{Num}, we discuss the numerical outocmes in detail. Further in section \ref{con}, we conclude our discussion along with mentioning the immediate future prospects of the present workdone in this paper. Finally, in Appendix \ref{bogoCoeff} and \ref{A1} we provide some of the necessary details which will be helpful to understand the content of this paper.
%the present work we work with this power spectrum and place constraints on the model-parameters in order to obviate the PBH Overproduction problem. %

\section{Basics of EFT within Bouncing Cosmology}\label{EFTBasic}
Notably, with the full diffeomorphism symmetry, such contribution becomes a scalar if we characterize the formulation in terms of scalar field degrees of freedom:
\bea
x^\nu\longrightarrow x^\nu+\xi^\nu(t,{\bf x}) \;\forall \nu=0,1,2,3.
\eea
The diffeomorphism parameter is denoted by $\xi^{\nu}(t,{\bf x})$ in this framework. One way to communicate these precise changes is as follows:
\begin{widetext}
    \bea
	t&&\longrightarrow t,~x^{i}\longrightarrow   x^{i}+\xi^{i}(t,{\bf x})~~~\forall~ i=1,2,3
~~~\delta\phi\longrightarrow \delta\phi,\\	
	t&&\longrightarrow t+ \xi^{0}(t,{\bf x}),~x^{i}\longrightarrow x^{i}~~~\forall~ i=1,2,3
	~~~\delta\phi\longrightarrow \delta\phi +\dot{\phi}_{0}(t)\xi^{0}(t,{\bf x}).\quad\quad
\eea
\end{widetext}
We define $\xi^{0}(t,{\bf x})$ and $\xi^{i}(t,{\bf x})\forall i = 1,2,3$, respectively, as the spatial and temporal diffeomorphism parameters. Under such circumstances, we use the gravitational gauge $\phi(t,{\bf x})=\phi_{0}(t)$. This homogeneous, isotropic, and spatially flat FLRW space-time is mirrored in $\phi_0(t)$, and it embeds the background time-dependent scalar field. To top it off, \bea \delta \phi(t,{\bf x})=0,\eea in this gauge choice.

The background geometry is described by the FLRW metric, which is described by the infinitesimal line element:
\bea
ds^2 = a^2(\tau)(-d\tau^2+d{\bf x}^2),
\eea

where the scale factor is represented by the scale factor $a(\tau)$, whose solution are given by the following expression.
\bea \label{scalef} a(\tau) =
\left\{
	\begin{array}{ll}
		a_0\left(\displaystyle\frac{\tau}{\tau_0}\right)^{\frac{1}{\epsilon-1}}  & \mbox{for } {\bf Case-I} \\
		a_0\left[1+\left(\displaystyle\frac{\tau}{\tau_0}\right)^2\right]^{\frac{1}{2(\epsilon-1)}} & \mbox{for } {\bf Case-II}
	\end{array}
\right.\eea
The following physical scenarios are characterized by the structure of the scale factors:
\begin{enumerate}
  \item The scale factor's power law solution is shown here as {\bf Case-I} \cite{Khoury:2001wf,Khoury:2001zk,Khoury:2001bz,Buchbinder:2007ad,Lehners:2007ac,Lehners:2008vx,Raveendran:2018yyh,Brandenberger:2012zb,Raveendran:2017vfx,Chowdhury:2015cma,Cai:2011tc,Brandenberger:2016vhg,Boyle:2004gv,Wands:1998yp,Peter:2002cn,Allen:2004vz,Martin:2003sf,Papanikolaou:2024fzf,Raveendran:2023auh,Raveendran:2019idj,Brustein:1998kq,Starobinsky:1980te,Mukhanov:1991zn,Brandenberger:1993ef,Novello:2008ra,Lilley:2015ksa,Battefeld:2014uga,Peter:2008qz,Biswas:2005qr,Bamba:2013fha,Nojiri:2014zqa,Bajardi:2020fxh,Bhargava:2020fhl,Cai:2009in,Cai:2012ag,Shtanov:2002mb,Ilyas:2020qja,Ilyas:2020zcb,Zhu:2021whu,Banerjee:2016hom,Saridakis:2018fth,Barca:2021qdn,Wilson-Ewing:2012lmx,K:2023gsi,Agullo:2020cvg,Agullo:2020fbw,Agullo:2020wur,Chowdhury:2018blx,Chowdhury:2016aet,Nandi:2019xag,Raveendran:2018why,Raveendran:2018yyh,Stargen:2016cft,Sriramkumar:2015yza,Banerjee:2022gpy,Paul:2022mup,Odintsov:2021yva,Banerjee:2020uil,Das:2017jrl,Pan:2024ydt,Colas:2024xjy,Piao:2003zm,Cai:2017dyi,Cai:2017pga,Cai:2015nya,Cai:2019hge}. In the quasi-de Sitter phase of inflation, the expansion is defined by a parameter $\epsilon<1$. Consider instances where $1<\epsilon<3$, where the matter contracting phase solution from {\bf Case I} is represented by $\epsilon=3/2$. From this power law solution, there is another fascinating scenario where $\epsilon>3$, which in this case reflects the ekpyrotic contracting phase. Moreover, it is crucial to remember that the scale factor at the conformal time scale $\tau=\tau_0$ is represented by the symbol $a(\tau_0)=a_0$, which establishes the reference for the {\bf Case-I}. The corresponding reference scale factors for the corresponding conformal time scales, $\tau_0=\tau_i$, $\tau_0=\tau_{ec}$, and $\tau_0=\tau_{mc}$, are given by $a_0=a_i$, $a_0=a_{ec}$, and $a_0=a_{mc}$, respectively, when we attempt to explain inflation, ekpyrotic contraction, and matter contraction scenarios with the above form of power law parametrization of the scale factor.

  \item In this instance, the scale factor describing the bouncing solution is represented as {\bf Case-II} \cite{Khoury:2001wf,Khoury:2001zk,Khoury:2001bz,Buchbinder:2007ad,Lehners:2007ac,Lehners:2008vx,Raveendran:2018yyh,Brandenberger:2012zb,Raveendran:2017vfx,Chowdhury:2015cma,Cai:2011tc,Brandenberger:2016vhg,Boyle:2004gv,Wands:1998yp,Peter:2002cn,Allen:2004vz,Martin:2003sf,Papanikolaou:2024fzf,Raveendran:2023auh,Raveendran:2019idj,Brustein:1998kq,Starobinsky:1980te,Mukhanov:1991zn,Brandenberger:1993ef,Novello:2008ra,Lilley:2015ksa,Battefeld:2014uga,Peter:2008qz,Biswas:2005qr,Bamba:2013fha,Nojiri:2014zqa,Bajardi:2020fxh,Bhargava:2020fhl,Cai:2009in,Cai:2012ag,Shtanov:2002mb,Ilyas:2020qja,Ilyas:2020zcb,Zhu:2021whu,Banerjee:2016hom,Saridakis:2018fth,Barca:2021qdn,Wilson-Ewing:2012lmx,K:2023gsi,Agullo:2020cvg,Agullo:2020fbw,Agullo:2020wur,Chowdhury:2018blx,Chowdhury:2016aet,Nandi:2019xag,Raveendran:2018why,Raveendran:2018yyh,Stargen:2016cft,Sriramkumar:2015yza,Banerjee:2022gpy,Paul:2022mup,Odintsov:2021yva,Banerjee:2020uil,Das:2017jrl,Pan:2024ydt,Colas:2024xjy,Piao:2003zm,Cai:2017dyi,Cai:2017pga,Cai:2015nya,Cai:2019hge}.  In particular, the matter bounce solution in this case corresponds to $\epsilon=3/2$. In contrast, the ekpyrotic bounce solution in this linked issue is physically represented by the circumstance that characterizes $\epsilon>3$. In light of this explanation, it is also required to mention that the reference scale for this particular {\bf Case-II} is established by $a(\tau_0)=a_0$, which represents the appropriate scale factor at the conformal time scale, $\tau=\tau_0$. The equivalent characteristic scale factors for the ekpyrotic and matter bounce situations are $a_0=a_{eb}$ and $a_0=a_{mb}$, respectively, when we use this scale factor solution to try to explain them.

\end{enumerate}
The following simplified expression can describe the generic form of the EFT action:
\begin{widetext}
   \bea
S&=&\int d^4x\sqrt{-g}\bigg[\frac{M_{pl}^2}{2}R+M_{pl}^2\dot{H}g^{00}-M_{pl}^2(3{\cal H}^{2}+\dot{H})+{\cal F}(\delta g^{00},\delta K^{\mu\nu},...)\bigg]
\eea 
\end{widetext}
Not to be overlooked, the final term in the previously stated expression ${\cal F}(\delta g^{00},\delta K^{\mu\nu},\cdots)$ quantifies all contributions from the small perturbation as they are described in Case-I and Case-II scenarios and can be expressed in the following simplified mathematical form:
\begin{widetext}
  \bea {\cal F}(\delta g^{00},\delta K^{\mu\nu},...):&=&\bigg[\frac{M^{4}_{2}(t)}{2!}\left(\delta g^{00}\right)^2+\frac{M^{4}_{3}(t)}{3!}\left(\delta g^{00}\right)^3-\frac{\bar{M}^{3}_{1}(t)}{2}\left(\delta g^{00}\right)\delta K^{\mu}_{\mu}-\frac{\bar{M}^{2}_{2}(t)}{2}(\delta K^{\mu}_{\mu})^2-\frac{\bar{M}^{2}_{3}(t)}{2}\delta K^{\mu}_{\nu}\delta K^{\nu}_{\mu}\bigg].\quad\eea  
\end{widetext}
    In this expression, $M_1(t)$, $M_3(t)$,  $\bar{M}_1(t)$,  $\bar{M}_2(t)$ and $\bar{M}_3(t)$ play the role of Wilson coefficients which one needs to fix from the analysis presented in the work. In addition, the final structure of the representative EFT action is generated by the polynomial powers of the fluctuation in the extrinsic curvature computed at constant time slice ($\delta K_{\mu\nu}$), which can be expressed as, $\delta K_{\mu\nu}=\left(K_{\mu\nu}-a^2Hh_{\mu\nu}\right)$, where the extrinsic curvature ($K_{\mu\nu}$), unit normal vector ($n_{\mu}$), and the induced metric ($h_{\mu\nu}$).

Under the temporal diffeomorphism symmetry, the Goldstone mode $(\pi(t,{\bf x}))$ varies, and we have,
$\pi(t,{\bf x})\longrightarrow \Tilde{\pi}(t,{\bf x})=\pi(t,{\bf x})-\xi^{0}(t,{\bf x})$. The local parameter is $\xi^0(t,{\bf x})$ in this case. This paper compares the role of these Goldstone modes to that of the scalar modes in cosmic perturbation. This indicates that the fixing criteria for the relevant unitary gauge is $\pi(t,x) = 0$. Here $\Tilde{\pi}(t,{\bf x})=-\xi^0(t,{\bf x})$ is suggested by this. With the perturbation indicated by $\delta g^{00}$ and $\bar{g}^{00}=-1$, we may express the temporal component of the metric after perturbation as $g^{00}=\bar{g}^{00}+\delta g^{00}$. The mixing contributions from gravity and the Goldstone modes may be readily disregarded in this limit. Important to keep in mind is that, in the decoupling limit, it may be simple to ignore the mixing term $E_{mix}=\sqrt{\dot{H}},$ which is located above the typical energy scale.

In the decoupling limit, the second-order perturbed action for the Goldstone modes can now be expressed in the following fashion:
\bea
S_{(2)}&&\approx%\int d^4x \;a^3\bigg[-M_{pl}^2\dot{H}\bigg(\dot{\pi}^2-\frac{1}{a^2}(\partial_i\pi)^2\bigg)\bigg]\nn\\&&%
\int d^4x\; a^3\bigg(\frac{-M_{pl}^2\dot{H}}{c_s^2}\bigg)\bigg(\dot{\pi}^2-c_s^2\frac{(\partial_i\pi)^2}{a^2}\bigg)\label{action}
\eea
Using the linear relation, $\zeta(t,{\bf x})=-H\pi(t,{\bf x}),$ which identifies comoving curvature perturbation with Goldstone modes along with proper Fourier transform ansatz equation(\ref{action}) gives:
\bea
S_{(2)}=\int \frac{d^3k}{(2\pi)^3}d\tau \;a^2 \bigg(\frac{M^2_pl\epsilon}{c_s^2}\bigg)(\zeta_k^{'2} -k^2c_s^2\zeta_k^2)
\eea
Varying the action would give us the Mukhanov-Sasaki equation which is represented in the Fourier space as:
\bea\label{MS Eq}
\zeta_{\bf k}^{''}(\tau)+2\frac{z^{'}(\tau)}{z(\tau)}\zeta_{\bf k}^{'}(\tau)+c_s^2k^2\zeta_{\bf k}(\tau)=0
\eea
Here, we introduce a conformal time-dependent new variable $z$, which is defined as, \bea z=a\sqrt{2\epsilon}/c_s\quad {\rm where}\quad \frac{z^{''}(\tau)}{z(\tau)}=\frac{1}{\tau^2}\left(\nu^2-\frac{1}{4}\right).\eea Now the definition of the effective sound speed in terms of the EFT Wilson coefficient is,
\bea c_s\equiv 1/\sqrt{1-\frac{2M_2^4}{\dot{H}M_{pl}^2}}.\eea
Once the coupling parameter $M_2$ is fixed, it will automatically constrain the EFT sound-speed parameter $c_s$ within a preferred window that respects both cosmological observations \cite{Planck:2018jri} as well as the causality-unitarity bound. 

The solutions for the Mukhanov-Sasaki equation (\ref{MS Eq}) for different phases are:
%The curvature perturbation for the {\bf Phase-I (Contraction)} with the Bunch-Davies initial condition has the following form:%

\begin{widetext}

\bea &&\underline{\rm \bf Phase-I~ (Contraction):}\;\nonumber\\ \label{contraction}
{\bf \zeta}_{\bf C}&=&\frac{2^{\nu-\frac{3}{2}} c_s (-k c_s \tau )^{\frac{3}{2}-\nu}}{ia_0\tau \sqrt{2 \epsilon_*}(k c_s)^{\frac{3}{2}}\sqrt{2} M_{pl}}\left(\frac{\tau}{\tau_0}\right)^{-\frac{1}{(\epsilon-1)}}\sqrt{\left(\frac{\epsilon_*}{\epsilon_c}\right)}\Bigg|\frac{\Gamma(\nu)}{\Gamma(\frac{3}{2})}\Bigg | (1+i k c_s\tau) e^{-i\left(k c_s\tau+\frac{\pi}{2}\left(\nu+\frac{1}{2}\right)\right)}.\quad\quad
\\ &&\underline{\rm \bf Phase-II~ (Bounce):}\;\nonumber\\\label{bounce}
{\bf \zeta}_{\bf B}&=&\frac{2^{\nu-\frac{3}{2}} c_s (-k c_s \tau )^{\frac{3}{2}-\nu}}{ia_0\tau \sqrt{2 \epsilon_*}(k c_s)^{\frac{3}{2}}\sqrt{2}  M_{pl}}\left[1+\left(\frac{\tau}{\tau_0}\right)^2\right]^{-\frac{1}{2(\epsilon-1)}}\Bigg|\frac{\Gamma(\nu)}{\Gamma(\frac{3}{2})}\Bigg |\sqrt{\left(\frac{\epsilon_*}{\epsilon_b}\right)}(1+i k c_s\tau) e^{-i\left(k c_s\tau+\frac{\pi}{2}\left(\nu+\frac{1}{2}\right)\right)}.\quad\quad
\\ &&\underline{\rm \bf Phase-III~ (First~Slow-Roll):}\;\nonumber\\ \label{SRI-curvature}
{\bf \zeta}_{\bf SRI}&=&\frac{2^{\nu-\frac{3}{2}} c_s H (-k c_s \tau )^{\frac{3}{2}-\nu}}{i \sqrt{2 \epsilon_*}(k c_s)^{\frac{3}{2}}\sqrt{2} M_{pl}}\Bigg|\frac{\Gamma(\nu)}{\Gamma(\frac{3}{2})}\Bigg |(1+i k c_s\tau) e^{-i\left(k c_s\tau+\frac{\pi}{2}\left(\nu+\frac{1}{2}\right)\right)}.
\\ &&\underline{\rm \bf Phase-IV~ (Ultra~Slow-Roll):}\;\nonumber\\\label{USR}
{\bf \zeta}_{\bf USR}&=&\frac{2^{\nu-\frac{3}{2}} c_s   H (-k c_s \tau )^{\frac{3}{2}-\nu}}{i \sqrt{2 \epsilon_*}(k c_s)^{\frac{3}{2}}\sqrt{2} M_{pl}}\bigg(\frac{\tau_s}{\tau}\bigg)^3\Bigg|\frac{\Gamma(\nu)}{\Gamma(\frac{3}{2})}\Bigg |\Bigg\{\alpha_2 (1+i k c_s\tau) e^{-i\left(k c_s\tau+\frac{\pi}{2}\left(\nu+\frac{1}{2}\right)\right)}-\beta_2(1-i k c_s \tau)e^{i\left(k c_s\tau+\frac{\pi}{2}\left(\nu+\frac{1}{2}\right)\right)}\Bigg\}.\quad\quad
\\ &&\underline{\rm \bf Phase-V~ (Second~Slow-Roll):}\;\nonumber\\\label{SRII-curvature}
{\bf \zeta}_{\bf SRII}&=&\frac{2^{\nu-\frac{3}{2}} c_s   H (-k c_s \tau )^{\frac{3}{2}-\nu}}{i \sqrt{2 \epsilon_*}(k c_s)^{\frac{3}{2}}\sqrt{2} M_p}\left(\frac{\tau_s}{\tau_e}\right)^3\Bigg|\frac{\Gamma(\nu)}{\Gamma(\frac{3}{2})}\Bigg |\Bigg\{\alpha_3 (1+i k c_s\tau) e^{-i\left (kc_s\tau+\frac{\pi}{2}(\nu+\frac{1}{2})\right )}-\beta_3(1-i k c_s \tau)e^{i\left(k c_s\tau+\frac{\pi}{2}(\nu+\frac{1}{2})\right)}\Bigg\}.\quad\quad
\eea
\end{widetext}
Here it is important to note that in the first three phases, i.e., contraction, bouncing, and the first slow-roll phase Bunch Davies initial vacuum has been chosen for the computational purpose. Due to having sharp transitions from SRI to USR and USR to SRII in the fourth and fifth phases the internal structure of the vacuum changes which is characterized by the Bogoliubov coefficients $\alpha_2$, $\beta_2$, $\alpha_3$ and $\beta_3$, which are given in the appendix(\ref{bogoCoeff}). The specific structure of these Bogoliubov coefficients can be expressed in terms of the Bunch Davies initial condition by making use of the continuity and differentiability of the obtained mode solutions from the Mukhanov-Sasaki equation.

\begin{figure*}[ht!]
    \centering
    \subfigure[]{
    \includegraphics[height=8cm,width=8.5cm]{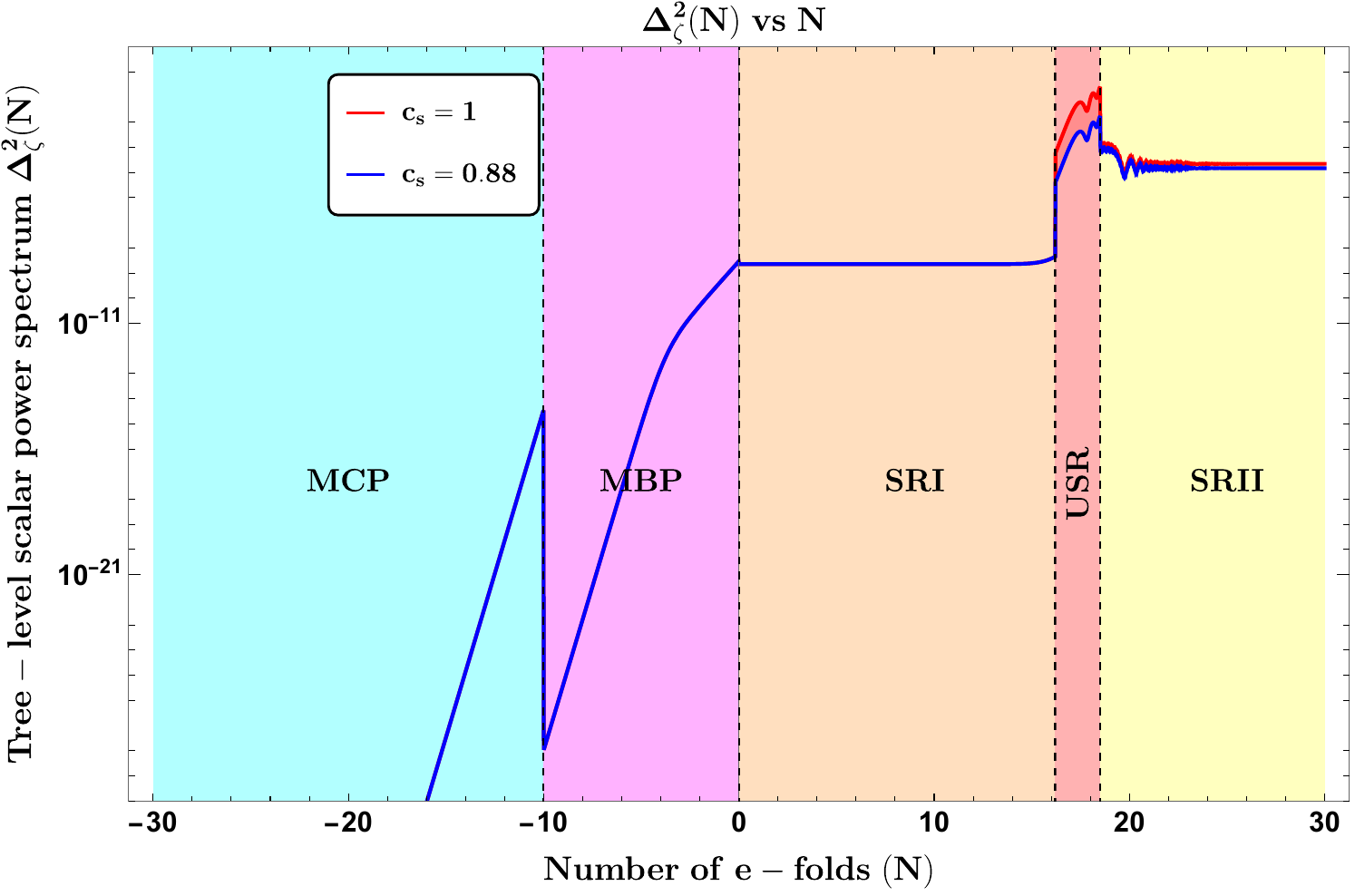}\label{g1}
    }
    \subfigure[]{
    \includegraphics[height=8cm,width=8.5cm]{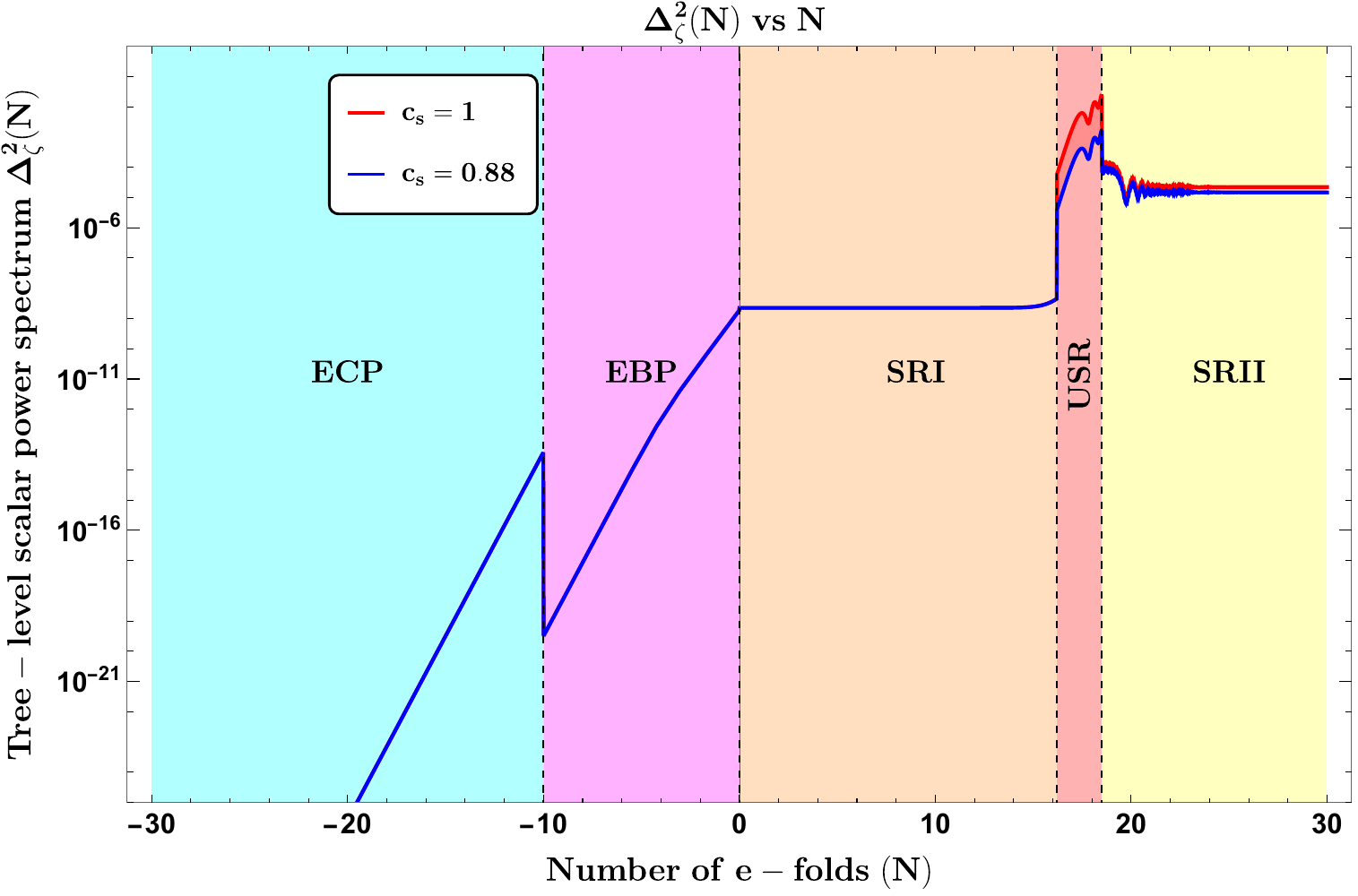}\label{g2}
    }
    \caption{Plot of tree-level power spectrum as a function of number of e-foldings. On left is matter contraction and bounce with the slow-roll parameter fixed to $\epsilon = 3/2$. On the right is the ekpyrotic contraction and bounce phases with the slow-roll parameter set to $\epsilon=7/2$. }
    \label{treeLevelPspec}
\end{figure*}

\section{Tree level power Specturm}\label{TreePS}

The comoving curvature perturbation is identified as occurring on the late time scale, $\tau\rightarrow 0$. With this knowledge, the tree-level contribution to the comoving curvature perturbation's two-point cosmic correlation function is expressed as follows:
\bea \langle \hat{\zeta}_{\bf k}\hat{\zeta}_{{\bf k}^{'}}\rangle_{{\bf Tree}} &=&\lim_{\tau\rightarrow 0}\langle \hat{\zeta}_{\bf k}(\tau)\hat{\zeta}_{{\bf k}^{'}}(\tau)\rangle_{{\bf Tree}}\nonumber\\
&=&(2\pi)^{3}\;\delta^{3}\left({\bf k}+{\bf k}^{'}\right)P^{\bf Tree}_{\zeta}(k),\quad\eea
In Fourier space, the dimensionful power spectrum is represented by the notation $P^{\bf Tree}_{\zeta}(k)$. It may be written as follows:
\bea P^{\bf Tree}_{\zeta}(k)=\langle \hat{\zeta}_{\bf k}\hat{\zeta}_{-{\bf k}}\rangle_{(0,0)}=|{\zeta}_{\bf k}(\tau)|^{2}_{\tau\rightarrow 0}.\quad\eea
Nonetheless, it becomes necessary to employ the dimensionless form of the power spectrum in Fourier space for practical reasons and in relation to cosmic measurements. This form is expressed as follows:
\bea \label{treex} \Delta^{2}_{\zeta,{\bf Tree}}(k)=\frac{k^{3}}{2\pi^{2}}P^{\bf Tree}_{\zeta}(k)=\frac{k^{3}}{2\pi^{2}}|{\zeta}_{\bf k}(\tau)|^{2}_{\tau\rightarrow 0}.\eea
The final tree-level power spectrum comes out to be:
\begin{widetext}
  \bea
\Delta^2_{\zeta,{\rm\bf Tree-Total}}(k)&=&\bigg[\Delta^{2}_{\zeta,{\bf Tree}}(k)\bigg]_{\bf CONT}+\bigg[\Delta^{2}_{\zeta,{\bf Tree}}(k)\bigg]_{\bf BOUNCE}+\bigg[\Delta^{2}_{\zeta,{\bf Tree}}(k)\bigg]_{\bf SRI}+\bigg[\Delta^{2}_{\zeta,{\bf Tree}}(k)\bigg]_{\bf USR}+\bigg[\Delta^{2}_{\zeta,{\bf Tree}}(k)\bigg]_{\bf SRII}\nonumber\\
&=&\Delta^2_{\zeta,{\rm\bf SRI}}(k_*
)\times\bigg[1+\bigg(\frac{\epsilon_*}{\epsilon_c}\bigg)\times\bigg(\frac{k}{k_*} \bigg)^{\frac{2\epsilon_c}{\epsilon_c-1}}+\bigg(\frac{\epsilon_*}{\epsilon_b}\bigg)\times\bigg(\frac{k}{k_*}\bigg)^2
\times\bigg[1+\bigg(\frac{k_*}{k}\bigg)^2\bigg]^{-\frac{1}{\epsilon_b-1}}\nonumber\\&&\quad\quad\quad\quad\quad\quad\quad\quad+\Theta(k-k_s)\bigg(\frac{k}{k_*}\bigg)^6\times|\alpha_2-\beta_2|^2+\Theta(k-k_e)\bigg(\frac{k_e}{k_s}\bigg)^6\times|\alpha_3-\beta_3|^2\bigg].
\eea  
\end{widetext}

Here, the two Heaviside Theta functions characterize the SRI to USR and USR to SRII sharp transitions.  The detailed derivation for the above expression can be found in the ref \cite{Choudhury:2024dei}. The plot of the tree level power spectrum as a function of the number of e-foldings is given in fig \ref{treeLevelPspec}. ae have separately considered two possibilities of having matter contraction with bounce (for slow-roll parameter $\epsilon=7/2$) and ekpyrotic contraction with bounce (for slow-roll parameter $\epsilon=3/2$) in figures \ref{g1} and \ref{g2} respectively. Comparing both of these plots, one can clearly visualize that due to having two different values of slow-roll parameters in the matter and ekpyrotic contraction-bounce scenarios the dynamical features are completely different in these two phases and this is represented via the computed perturbations modes from the Mukhanov Sasaki equation as mentioned in the previous section. Such differences are directly related to the tree-level power spectrum as we can see that the matter contracting phase (MCP) is different from the ekpyrotic contracting phase (ECP) and the matter bouncing phase (MBP) is different from the ekpyrotic bouncing phase (EBP). The rest of the behaviour appeasing for SRI, USR, and SRII phases in figures \ref{g1} and \ref{g2} becomes exactly identical as the behaviour of the slow-roll parameter $\epsilon$ are identical in each of the individual phases for both of these plots. Additionally, it is important to note that from the figures \ref{g1} and \ref{g2} the span of the USR phase in both of these cases in terms of the number of e-foldings is very small i.e. $N_{\rm USR}\sim 2$. This is extremely important because it helps to maintain the perturbative approximations during performing the computation. Moreover, it is also important to highlight that due to the appearance of new Bogoliubov coefficients $\alpha_2$, $\beta_2$, $\alpha_3$ and $\beta_3$, (see appendix(\ref{bogoCoeff})), which captures the impacts of changing vacuum in USR and SRII phases compared to the Bunch Davies vacuum as used to describe the features in the contracting, bouncing and SRI phases, small amplitude oscillating effects are incorporated in the tree-level power spectrum in the USR and SRII phases respectively. Furthermore, after comparing figures \ref{g1} and \ref{g2} one can clearly visualize that in the USR and SRII phases small difference in the spectrum with respect to the outcomes obtained from two different values of the EFT sound-speed parameter $c_s=0.88$ and $c_s=1$. In other phases, i.e. contraction, bounce and SRI region such differences cannot be clearly visible in the corresponding scalar power spectrum.

\section{Regularized-Renormalized-Resummed loop corrected Power Spectrum}\label{RRR}

We shall next do a simple estimate of the impact of one-loop corrections on the power spectrum resulting from scalar perturbation modes. The following calculation will be performed assuming that the usual EFT action expands in third order due to the curvature perturbation:
\begin{widetext}
    \bea &&S^{(3)}_{\zeta}=\int d\tau\;{\cal L}_3=\int d\tau\;  d^3x\;  M^2_{ pl}a^2\; \bigg[\left(3\left(c^2_s-1\right)\epsilon+\epsilon^2-\frac{1}{2}\epsilon^3\right)\zeta^{'2}\zeta+\frac{\epsilon}{c^2_s}\bigg(\epsilon-2s+1-c^2_s\bigg)\left(\partial_i\zeta\right)^2\zeta\nonumber\\ 
&&\quad\quad\quad\quad\quad\quad\quad\quad\quad\quad\quad-\frac{2\epsilon}{c^2_s}\zeta^{'}\left(\partial_i\zeta\right)\left(\partial_i\partial^{-2}\left(\frac{\epsilon\zeta^{'}}{c^2_s}\right)\right)-\frac{1}{aH}\left(1-\frac{1}{c^2_{s}}\right)\epsilon \bigg(\zeta^{'3}+\zeta^{'}(\partial_{i}\zeta)^2\bigg)
     \nonumber\\
&& \quad\quad\quad\quad\quad\quad\quad\quad\quad\quad\quad+\frac{1}{2}\epsilon\zeta\left(\partial_i\partial_j\partial^{-2}\left(\frac{\epsilon\zeta^{'}}{c^2_s}\right)\right)^2
+\underbrace{\frac{1}{2c^2_s}\epsilon\partial_{\tau}\left(\frac{\eta}{c^2_s}\right)\zeta^{'}\zeta^{2}}_{\bf Most~dominant ~term~in~USR}+\cdots
  \bigg],\quad\quad\eea
\end{widetext}
Here the $\cdots$ represent the higher order suppressed contributions which can be neglected in the present part of the computation. The term that is highlighted has the highest contribution to the first, second, and USR regions, denoted as ${\cal O}(\epsilon^{3})$, ${\cal O}(\epsilon^{3})$, and ${\cal O}(\epsilon)$, respectively. It is also crucial to take into account that the term that is emphasized above has a little impact during the contraction phase and a significant suppression during the bounce phase. The final five contributions are operators with a strong Planck suppression, resulting in a negligible adjustment to the one-loop power spectrum for each of the five phases that were previously described. It is also important to take into account the fact that the term that was emphasized above has a negligible contribution during the contraction phase and a significant suppression during the bounce phase. 

The most important highlighted term that results from the EFT framework that we have selected for our current research will be the subject of our explicit analysis of each term's contributions as they occur in the third-order action in this paragraph. We apply the widely recognized in-in formalism to this aim.  This suggests that the next quantum operator's two-point correlation function at the late time scale, $\tau\rightarrow 0$, might be expressed as follows:
\begin{widetext}
     \bea \label{aamx}\langle\hat{\zeta}_{\bf p}\hat{\zeta}_{-{\bf p}}\rangle:&=&
    \lim_{\tau\rightarrow 0}\left\langle\bigg[\overline{T}\exp\bigg(i\int^{\tau}_{-\infty(1-i\epsilon)}d\tau^{'}\;H_{\rm int}(\tau^{'})\bigg)\bigg]\;\;\hat{\zeta}_{\bf p}(\tau)\hat{\zeta}_{-{\bf p}}(\tau)
\;\;\bigg[{T}\exp\bigg(-i\int^{\tau}_{-\infty(1+i\epsilon)}d\tau^{''}\;H_{\rm int}(\tau^{''})\bigg)\bigg]\right\rangle\nonumber\\
&=& \langle\hat{\zeta}_{\bf p}\hat{\zeta}_{-{\bf p}}\rangle_{(0,0)}+=\langle\hat{\zeta}_{\bf p}\hat{\zeta}_{-{\bf p}}\rangle_{(0,1)}+\langle\hat{\zeta}_{\bf p}\hat{\zeta}_{-{\bf p}}\rangle^{\dagger}_{(0,1)}+\langle\hat{\zeta}_{\bf p}\hat{\zeta}_{-{\bf p}}\rangle_{(0,2)}+\langle\hat{\zeta}_{\bf p}\hat{\zeta}_{-{\bf p}}\rangle^{\dagger}_{(0,2)}+\langle\hat{\zeta}_{\bf p}\hat{\zeta}_{-{\bf p}}\rangle_{(1,1)},\eea
\end{widetext}
where we made explicit use of the Legendre transformation, $H_{\rm int}=-{\cal L}_3$, which links the interaction Hamiltonian to the Lagrangian density that characterizes the third-order perturbation. We must assess the explicit contributions that are shown in the one-loop level and tree level results of the two-point correlation function of the scalar modes, which are attached below. These contributions are present for the five successive phases that were previously mentioned:
\begin{widetext}
    \bea
     &&\label{a0}\langle\hat{\zeta}_{\bf p}\hat{\zeta}_{-{\bf p}}\rangle_{(0,0)}=\lim_{\tau\rightarrow 0}\langle \hat{\zeta}_{\bf p}(\tau)\hat{\zeta}_{-{\bf p}}(\tau)\rangle,\\
    &&\label{a1}\langle\hat{\zeta}_{\bf p}\hat{\zeta}_{-{\bf p}}\rangle_{(0,1)}=-i\lim_{\tau\rightarrow 0}\int^{\tau}_{-\infty}d\tau_1\;\langle \hat{\zeta}_{\bf p}(\tau)\hat{\zeta}_{-{\bf p}}(\tau)H_{\rm int}(\tau_1)\rangle,\\
 &&\label{a2}\langle\hat{\zeta}_{\bf p}\hat{\zeta}_{-{\bf p}}\rangle^{\dagger}_{(0,1)}=-i\lim_{\tau\rightarrow 0}\int^{\tau}_{-\infty}d\tau_1\;\langle \hat{\zeta}_{\bf p}(\tau)\hat{\zeta}_{-{\bf p}}(\tau)H_{\rm int}(\tau_1)\rangle^{\dagger},\\
 &&\label{a3}\langle\hat{\zeta}_{\bf p}\hat{\zeta}_{-{\bf p}}\rangle_{(0,2)}=\lim_{\tau\rightarrow 0}\int^{\tau}_{-\infty}d\tau_1\;\int^{\tau}_{-\infty}d\tau_2\;\langle \hat{\zeta}_{\bf p}(\tau)\hat{\zeta}_{-{\bf p}}(\tau)H_{\rm int}(\tau_1)H_{\rm int}(\tau_2)\rangle,\\
 &&\label{a4}\langle\hat{\zeta}_{\bf p}\hat{\zeta}_{-{\bf p}}\rangle^{\dagger}_{(0,2)}=\lim_{\tau\rightarrow 0}\int^{\tau}_{-\infty}d\tau_1\;\int^{\tau}_{-\infty}d\tau_2\;\langle \hat{\zeta}_{\bf p}(\tau)\hat{\zeta}_{-{\bf p}}(\tau)H_{\rm int}(\tau_1)H_{\rm int}(\tau_2)\rangle^{\dagger},\\
  &&\label{a5}\langle\hat{\zeta}_{\bf p}\hat{\zeta}_{-{\bf p}}\rangle^{\dagger}_{(1,1)}=\lim_{\tau\rightarrow 0}\int^{\tau}_{-\infty}d\tau_1\;\int^{\tau}_{-\infty}d\tau_2\;\langle H_{\rm int}(\tau_1)\hat{\zeta}_{\bf p}(\tau)\hat{\zeta}_{-{\bf p}}(\tau)H_{\rm int}(\tau_2)\rangle^{\dagger}.\eea
\end{widetext}
The following equation may therefore be used to represent the entire one-loop regularized and renormalized one-loop adjusted power spectrum for scalar modes:
\begin{widetext}
     \bea \label{one-loopRR} \Delta^{2}_{\zeta, {\bf RR}}(k)&=&\bigg[\Delta^{2}_{\zeta,{\bf Tree}}(k)\bigg]_{\bf SRI}\times\bigg(1+\underbrace{\overline{{\bf W}}_{\bf C}+\overline{{\bf W}}_{\bf B}+\overline{{\bf W}}_{\bf SRI}+\overline{{\bf W}}_{\bf USR}+\overline{{\bf W}}_{\bf SRII}}_{\textbf{Regularized and Renormalized one-loop correction}}\bigg)\nonumber\\
   &=&\begin{tikzpicture}[baseline={([yshift=-3.5ex]current bounding box.center)},very thick]
  
    % Loop
  \def\radius{1}
  \scalebox{1}{\draw[red,very thick] (0,\radius) circle (\radius);
  \draw[red,very thick] (4.5*\radius,0) circle (\radius);}

  % External lines
  %\filldraw;
  \draw[black, very thick] (-4*\radius,0) -- 
  (-2.5*\radius,0);
  \node at (-2*\radius,0) {+};
  \draw[black, very thick] (-1.5*\radius,0) -- (0,0);
  \draw[blue,fill=blue] (0,0) circle (.5ex);
  \draw[black, very thick] (0,0)  -- (1.5*\radius,0);
  \node at (2*\radius,0) {+};
  \draw[black, very thick] (2.5*\radius,0) -- (3.5*\radius,0); 
  \draw[blue,fill=blue] (3.5*\radius,0) circle (.5ex);
  \draw[blue,fill=blue] (5.5*\radius,0) circle (.5ex);
  \draw[black, very thick] (5.5*\radius,0) -- (6.5*\radius,0);
\end{tikzpicture},\eea
\end{widetext}
where we have used cut-off regularization and late-time renormalization schemes to achieve the above-mentioned result. To know more about the details on this derivation see the ref \cite{Choudhury:2024dei}. Here $\overline{{\bf W}}_{\bf C}$, $\overline{{\bf W}}_{\bf B}$, $\overline{{\bf W}}_{\bf SRI}$, $\overline{{\bf W}}_{\bf USR}$, and $\overline{{\bf W}}_{\bf SRII}$ are represented by the following expressions:
\begin{widetext}
    \bea \label{t1}\overline{\bf W}_{\bf C}&=&-\frac{4}{3}\bigg[\Delta^{2}_{\zeta,{\bf Tree}}(k)\bigg]_{\bf SRI}\times\Bigg(1-\frac{2}{15\pi^2}\frac{1}{c^2_{s}k^2_c}\left(1-\frac{1}{c^2_{s}}\right)\epsilon_c\Bigg)\times \left(\frac{\epsilon_*}{\epsilon_c}\right)\times\bigg[\frac{1}{\delta_{\bf C}}\bigg\{\left(\frac{k_{b}}{k_*}\right)^{\delta_{\bf C}}-\left(\frac{k_{c}}{k_*}\right)^{\delta_{\bf C}}\bigg\}\nonumber\\
&&\quad\quad\quad\quad\quad\quad\quad\quad\quad\quad\quad\quad\quad\quad\quad\quad\quad\quad\quad\quad\quad\quad\quad\quad\quad+\frac{1}{\left(\delta_{\bf C}+2\right)}\bigg\{\left(\frac{k_{b}}{k_*}\right)^{\delta_{\bf C}+2}-\left(\frac{k_{c}}{k_*}\right)^{\delta_{\bf C}+2}\bigg\}\bigg],\\
\label{t2}\overline{\bf W}_{\bf B}&=&-\frac{4}{3}\bigg[\Delta^{2}_{\zeta,{\bf Tree}}(k)\bigg]_{\bf SRI}\times\Bigg(1-\frac{2}{15\pi^2}\frac{1}{c^2_{s}k^2_b}\left(1-\frac{1}{c^2_{s}}\right)\epsilon_b\Bigg)\times \left(\frac{\epsilon_*}{\epsilon_b}\right)\times\frac{1}{\delta_{\bf B} +2}\bigg[\, _2F_1\left(\frac{\delta_{\bf B}+2}{2},\frac{1}{\epsilon_b-1}-1;\frac{\delta_{\bf B}+4}{2};-1\right)\nonumber\\
 &&\quad\quad\quad\quad\quad\quad\quad\quad\quad\quad\quad\quad\quad\quad\nonumber\\
 &&\quad\quad\quad\quad\quad\quad\quad\quad\quad\quad\quad\quad\quad\quad\quad\quad-\left(\frac{k_{b}}{k_{*}}\right)^{\delta_{\bf B}+2}\, _2F_1\left(\frac{\delta_{\bf B}+2}{2},\frac{1}{\epsilon_b-1}-1;\frac{\delta_{\bf B}+4}{2};-\left(\frac{k_{b}}{k_{*}}\right)^{2}\right)\bigg],\\
 \label{t3}\overline{\bf W}_{\bf SRI}&=&-\frac{4}{3}\bigg[\Delta^{2}_{\zeta,{\bf Tree}}(k)\bigg]_{\bf SRI}\times\Bigg(1-\frac{2}{15\pi^2}\frac{1}{c^2_{s}k^2_*}\left(1-\frac{1}{c^2_{s}}\right)\epsilon_*\Bigg)\times\ln\left(\frac{k_s}{k_*}\right),\\
\label{t4}\overline{\bf W}_{\bf USR}&=&\frac{1}{4}\bigg[\Delta^{2}_{\zeta,{\bf Tree}}(k)\bigg]_{\bf SRI}\times\bigg[\bigg(\frac{\Delta\eta(\tau_{e})}{\tilde{c}^{4}_{s}}\bigg)^{2}{\bigg(\frac{k_{e}}{k_{s}}\bigg)^6} - \left(\frac{\Delta\eta(\tau_{s})}{\tilde{c}^{4}_{s}}\right)^{2}\bigg]\times\ln\left(\frac{k_e}{k_s}\right),\\
\label{t5}\overline{\bf W}_{\bf SRII}&=&\bigg[\Delta^{2}_{\zeta,{\bf Tree}}(k)\bigg]^2_{\bf SRI}\times\Bigg(1-\frac{2}{15\pi^2}\frac{1}{c^2_{s}k^2_*}\left(1-\frac{1}{c^2_{s}}\right)\epsilon_*\Bigg)\times\ln\left(\frac{k_{\rm end}}{k_e}\right).\eea
\end{widetext}
Here, $\delta_{\bf C}$ and $\delta_{\bf B}$ are defined as:
\bea \delta_{\bf C}:&=&\left(3-2\nu+\frac{2\epsilon_c}{\epsilon_c-1}\right),\\
\delta_{\bf B}:&=&\left(3-2\nu+\frac{2}{\epsilon_b-1}\right).\eea Here $\epsilon_c=3/2$ and $\epsilon_c>3$ for matter and ekpyrotic contraction respectively. Similarly, $\epsilon_b=3/2$ and $\epsilon_b>3$ for matter and ekpyrotic bounce. It is significant to note that the power spectrum at any scale $k$ during the SRI phase may be expressed as follows:
\bea \bigg[\Delta^{2}_{\zeta,{\bf Tree}}(k)\bigg]_{\bf SRI}
&=&\bigg[\Delta^{2}_{\zeta,{\bf Tree}}(k_*)\bigg]_{\bf SRI}\bigg(1+\bigg(\frac{k}{k_s}\bigg)^2\bigg),\quad\quad\eea
where at the CMB pivot scale $k_*$ we have following expression:
\bea\label{pivot} \bigg[\Delta^{2}_{\zeta,{\bf Tree}}(k_*)\bigg]_{\bf SRI}=\left(\frac{2^{2\nu-3}H^{2}}{8\pi^{2}M^{2}_{ pl}\epsilon c_s}\left|\frac{\Gamma(\nu)}{\Gamma\left(\frac{3}{2}\right)}\right|^2\right)_*.\eea
Further, we compute the expression for the all-loop renormalized result, which is given by:
\bea \Delta^{2}_{\zeta, {\bf RR}}(k)
&=&\bigg[\Delta_{\zeta,\textbf{Tree}}^{2}(k)\bigg]_{\textbf{SRI}}\nonumber\\
&&\quad\quad\times \Bigg(1+\sum_{\textbf{All even graphs G}}{\cal F}_{\bf G}\Bigg),\eea
where the explicit mathematical structure of the all-loop renormalized factor $\sum_{\textbf{All even graphs G}}{\cal F}_{\bf G}$ is given by the following expression:
\begin{widetext}
 \bea   &&\sum_{\textbf{All even graphs G}}{\cal F}_{\bf G}\nonumber\\
&&=-\left\{\frac{\bigg[  \Delta_{\zeta,\textbf{Tree}}^{2}(k)\bigg]_{\textbf{SRI}}}{\bigg[  \Delta_{\zeta,\textbf{Tree}}^{2}(k_{*})\bigg]_{\textbf{SRI}}}\right\}\times\bigg[\bigg(\overline{{\bf W}}^2_{\bf C,*}+\overline{{\bf W}}^2_{\bf B,*}+\overline{{\bf W}}^2_{\bf SRI,*}+\overline{{\bf W}}^2_{\bf USR,*}+\overline{{\bf W}}^2_{\bf SRII,*}\bigg)+\cdots\bigg]\nonumber\\
&&=\begin{tikzpicture}[baseline={([yshift=-5.5ex]current bounding box.center)},very thick]

      % Loop
  \def\radius{0.5}
  \scalebox{0.5}{
  \draw[red, ultra thick] (3*\radius,0) circle (\radius);
  \draw[red,ultra thick] (5*\radius,0) circle (\radius);
  \draw[red,ultra thick] (13*\radius,\radius) circle (\radius);
  \draw[red,ultra thick] (13*\radius,3*\radius) circle (\radius);
  \draw[red,ultra thick] (21*\radius,0) circle (\radius);}

  % External lines
  %\filldraw
  %\draw[black, very thick] (-2*\radius,0) -- (-1*\radius,0); 
  %\node at (-0.5*\radius,0) {+};
  \draw[black, very thick] (0,0) -- (\radius,0); 
  \draw[blue,fill=blue] (\radius,0) circle (.3ex);
  \draw[blue,fill=blue] (2*\radius,0) circle (.3ex);
  \draw[blue,fill=blue] (3*\radius,0) circle (.3ex);
  \draw[black, very thick] (3*\radius,0) -- (4*\radius,0);
  \node at (4.5*\radius,0) {+};
  \draw[black, very thick] (5*\radius,0) -- (6.5*\radius,0);
  \draw[blue,fill=blue] (6.5*\radius,0,0) circle (.3ex);
  \draw[blue,fill=blue] (6.5*\radius,\radius) circle (.3ex);
  \draw[black, very thick] (6.5*\radius,0,0) -- (8*\radius,0);
  \node at (8.5*\radius,0) {+};
  \draw[black, very thick] (9*\radius,0) -- (10*\radius,0);
  \draw[blue,fill=blue] (10*\radius,0,0) circle (.3ex);
  \draw[red, ultra thick] (10*\radius,0) -- (11*\radius,0);
  \draw[blue,fill=blue] (11*\radius,0,0) circle (.3ex);
  \draw[black, very thick] (11*\radius,0) -- (12*\radius,0);
  \node at (12.5*\radius,0) {+};
  \draw[black, very thick](13*\radius,0) --(14*\radius,0);
  \draw[blue,fill=blue](14*\radius,0,0) circle (.3ex);
  \draw[red,  thick](14*\radius,0) --(14.5*\radius,\radius);
  \draw[blue,fill=blue](14.5*\radius,\radius) circle (.3ex);
\draw[red, thick](14.5*\radius,\radius) --(15.5*\radius,\radius);
\draw[blue,fill=blue](15.5*\radius,\radius) circle (.3ex);
\draw[red, thick](14*\radius,0) -- (14.5*\radius,-\radius);\draw[blue,fill=blue](14.5*\radius,-\radius) circle (.3ex);
\draw[red,  thick](14.5*\radius,\radius) -- (14.5*\radius,-\radius);
\draw[red, thick](14.5*\radius,-\radius) -- (15.5*\radius,-\radius);
\draw[blue,fill=blue](15.5*\radius,-\radius) circle (.3ex);
\draw[blue,fill=blue](14.5*\radius,0) circle (.3ex);
\draw[red, thick](14.5*\radius,0) -- (15.5*\radius,0);
\draw[red,  thick](15.5*\radius,\radius) -- (15.5*\radius,-\radius);
\draw[blue,fill=blue](15.5*\radius,0) circle (.3ex);
\draw[red, thick](15.5*\radius,\radius) -- (16*\radius,0);
\draw[blue,fill=blue](16*\radius,0) circle (.3ex);
\draw[red, thick](16*\radius,0) -- (15.5*\radius,-\radius);
\draw[black, very thick](16*\radius,0) -- (17*\radius,0);
\node at (17.5*\radius,0) {+};
\draw[black, very thick](18*\radius,0) -- (19*\radius,0);
\draw[blue,fill=blue](19*\radius,0) circle (.3ex);
\draw[red, thick](19*\radius,0) -- (19.5*\radius,\radius);
\draw[red, thick](19*\radius,0) -- (19.5*\radius,-\radius);
\draw[blue,fill=blue](19.5*\radius,\radius) circle (.3ex);
 \draw[blue,fill=blue](19.5*\radius,-\radius) circle (.3ex); 
 \draw[red, thick](19.5*\radius,\radius) -- (20.5*\radius,\radius);
 \draw[red, thick](19.5*\radius,-\radius) -- (20.5*\radius,-\radius);
 \draw[blue,fill=blue](20.5*\radius,\radius) circle (.3ex);
 \draw[blue,fill=blue](20.5*\radius,-\radius) circle (.3ex);
 \draw[red, thick](20.5*\radius,\radius) -- (21*\radius,0);
 \draw[red, thick](20.5*\radius,-\radius) -- (21*\radius,0);
\draw[blue,fill=blue](21*\radius,0) circle (.3ex);
\draw[black,  thick](21*\radius,0) -- (22*\radius,0); 
\draw[red, thick](19.5*\radius,\radius) -- (20.5*\radius,-\radius);
\draw[red, thick](20.5*\radius,\radius) -- (19.5*\radius,-\radius);
\node at (22.5*\radius,0) {+};
\draw[black, thick](23*\radius,0) -- (24*\radius,0);
\draw[red, thick](24*\radius,0) -- (24.5*\radius,\radius);
\draw[red, thick](24*\radius,0) -- (24.5*\radius,-\radius);
\draw[blue,fill=blue](24*\radius,0) circle (.3ex);
\draw[blue,fill=blue](24.5*\radius,\radius) circle (.3ex);
\draw[blue,fill=blue](24.5*\radius,-\radius) circle (.3ex);
\draw[red, thick](24.5*\radius,\radius) -- (25.5*\radius,\radius);
\draw[red,  thick](24.5*\radius,-\radius) -- (25.5*\radius,-\radius);
\draw[red,  thick](25.5*\radius,\radius) -- (26*\radius,0);
\draw[red, thick](26*\radius,0) -- (25.5*\radius,-\radius);
\draw[blue,fill=blue](25.5*\radius,\radius) circle (.3ex);
\draw[blue,fill=blue](26*\radius,0) circle (.3ex);
\draw[blue,fill=blue](25.5*\radius,-\radius) circle (.3ex);
\draw[red,  thick](24.5*\radius,\radius) -- (24.5*\radius,-\radius);
\draw[red, thick](25*\radius,\radius) -- (25*\radius,-\radius);
\draw[red, thick](25.5*\radius,\radius) -- (25.5*\radius,-\radius);
\draw[blue,fill=blue](25*\radius,\radius) circle (.3ex);
\draw[blue,fill=blue](25*\radius,-\radius) circle (.3ex);
\draw[black, thick](26*\radius,0) -- (27*\radius,0); 
\node at (28*\radius,0) {+};
\node at (29*\radius,0) {...};

\end{tikzpicture}.\eea
\end{widetext}

%%%%%%%%%%%%%%%%%%%%%%%%%%%%%%%%%%%%%%%%%%%%%
\begin{figure*}[htb!]
    \centering
    \subfigure[]{
    \includegraphics[height=8cm,width=8.5cm]{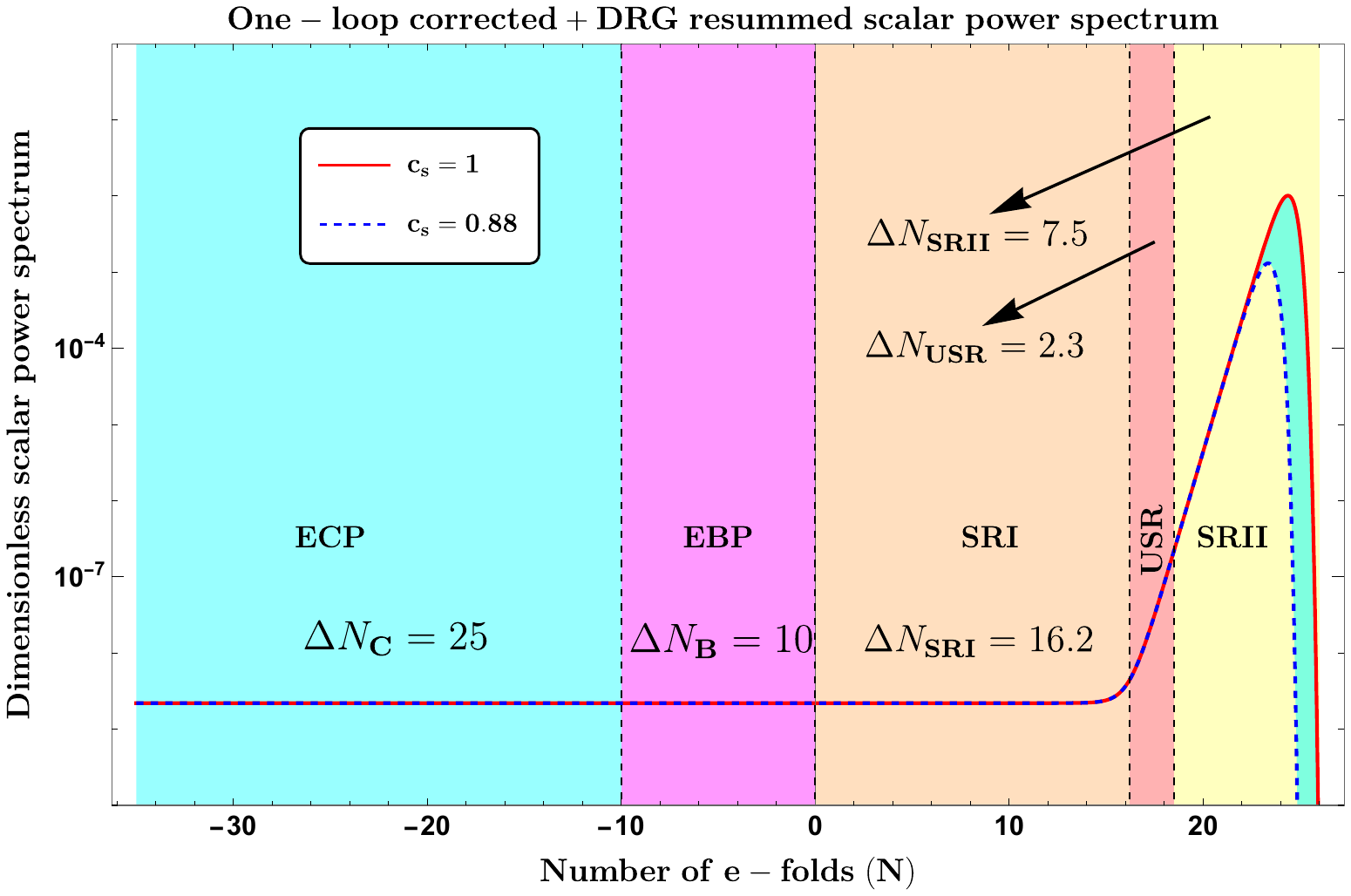}
    \label{RRR1}
    }
    \subfigure[]{
    \includegraphics[height=8cm,width=8.5cm]{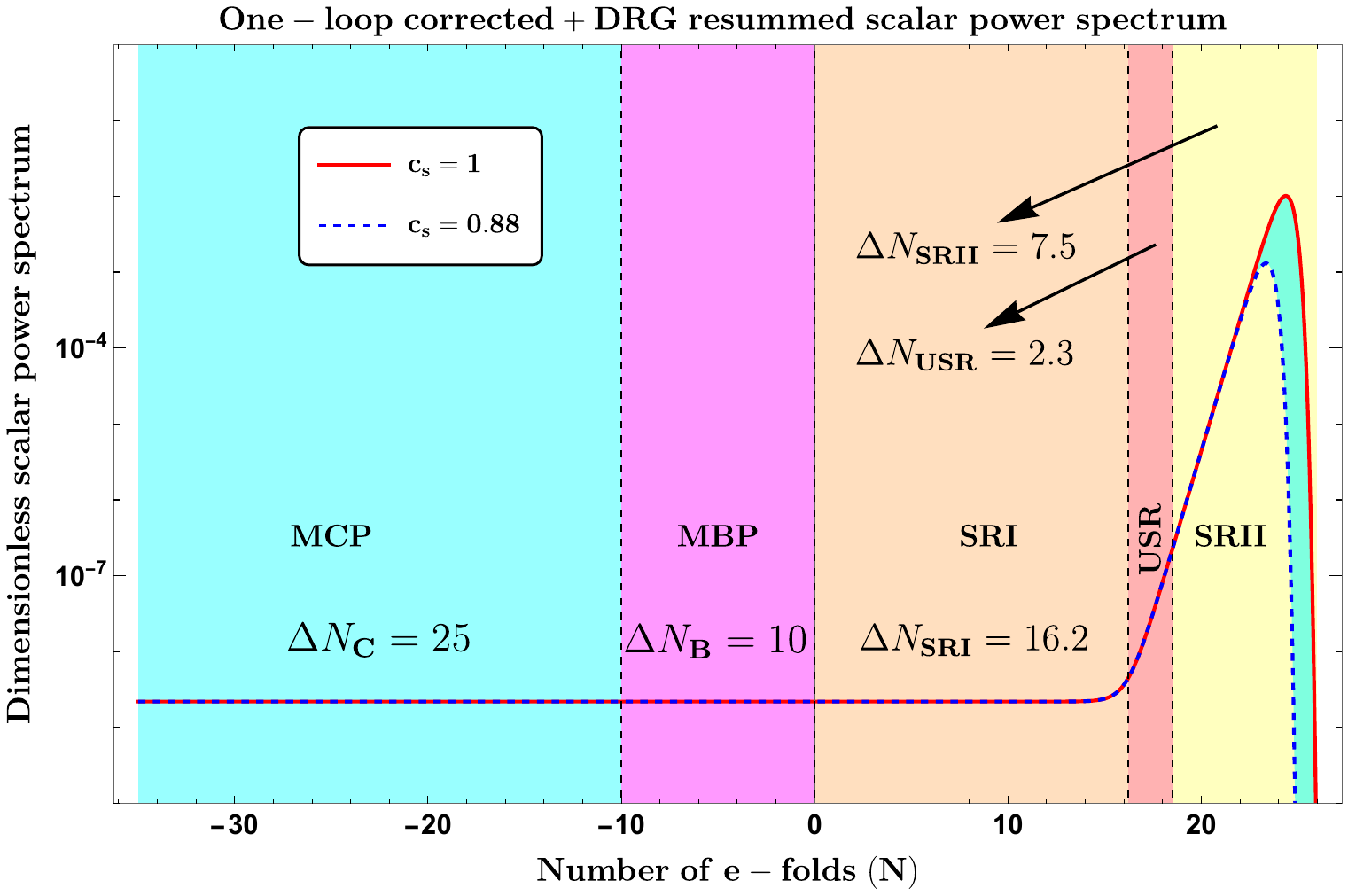}
    \label{RRR2}
    }
    \caption{Plots of regularized-renormalized-resummed scalar power spectrums as a function of the e-foldings N. The left has the ekpyrotic contraction and bounce phases with the slow-roll parameter $\epsilon=7/2$ fixed. The right one has the matter contraction and bounce phases with the slow-roll parameter $\epsilon=3/2$ fixed. The red and blue lines correspond to the effective sound speed $c_s=1,0.88$ respectively.}\label{RRRPspec}
\end{figure*}
To carry out the resummation and finally provide a finite result taking into account the contributions up to all even loop order, the total size of this component needs to be maintained inside the perturbative limit. This approach works nicely for our computing needs since we quote the result in this manner. Introducing the Dynamical Renormalization Group (DRG) approach \cite{Chen:2016nrs,Baumann:2019ghk,Boyanovsky:1998aa,Boyanovsky:2001ty,Boyanovsky:2003ui,Burgess:2015ajz,Burgess:2014eoa,Burgess:2009bs,Dias:2012qy,Chaykov:2022zro,Chaykov:2022pwd} is our major objective here, since it enables us to resum over all of the logarithmically divergent contributions in the present calculation. This resummed result is valid, too, for perturbative computations in all loop orders in which the quantum effects are well-captured. However, this can only occur if the resummed infinite series meets the stringent convergence conditions at super-horizon and horizon-crossing values. The aforementioned elements of the convergent series are all products of the cosmic perturbation of scalar modes hypothesis in all feasible loop orders. Generally speaking, the DRG process is the intrinsic mechanism that allows secular momentum-dependent contributions to the convergent infinite series at super-horizon and horizon crossing scales to be justified. When discussing and computing in the associated context, the DRG resummation is sometimes referred to as the resummation under the influence of exponentiation. Ultimately, the power spectrum of the scalar modes in their regularized-ronormalized-resummed (RRR) form may be represented by:
\begin{widetext}
    \bea \label{DRGc1} \Delta^{2}_{\zeta, {\bf RRR}}(k)
&=&\bigg[\Delta^{2}_{\zeta,{\bf Tree}}(k)\bigg]_{\bf SRI}\times\exp\bigg(\sum_{\textbf{All even graphs G}}{\cal F}_{\bf G}\bigg)\nonumber\\
&=&\bigg[\Delta^{2}_{\zeta,{\bf Tree}}(k)\bigg]_{\bf SRI}\\
&&\times\exp\left(\begin{tikzpicture}[baseline={([yshift=-5.5ex]current bounding box.center)},very thick]

      % Loop
  \def\radius{0.5}
  \scalebox{0.5}{
  \draw[red, ultra thick] (3*\radius,0) circle (\radius);
  \draw[red,ultra thick] (5*\radius,0) circle (\radius);
  \draw[red,ultra thick] (13*\radius,\radius) circle (\radius);
  \draw[red,ultra thick] (13*\radius,3*\radius) circle (\radius);
  \draw[red,ultra thick] (21*\radius,0) circle (\radius);}

  % External lines
  %\filldraw
  %\draw[black, very thick] (-2*\radius,0) -- (-1*\radius,0); 
  %\node at (-0.5*\radius,0) {+};
  \draw[black, very thick] (0,0) -- (\radius,0); 
  \draw[blue,fill=blue] (\radius,0) circle (.3ex);
  \draw[blue,fill=blue] (2*\radius,0) circle (.3ex);
  \draw[blue,fill=blue] (3*\radius,0) circle (.3ex);
  \draw[black, very thick] (3*\radius,0) -- (4*\radius,0);
  \node at (4.5*\radius,0) {+};
  \draw[black, very thick] (5*\radius,0) -- (6.5*\radius,0);
  \draw[blue,fill=blue] (6.5*\radius,0,0) circle (.3ex);
  \draw[blue,fill=blue] (6.5*\radius,\radius) circle (.3ex);
  \draw[black, very thick] (6.5*\radius,0,0) -- (8*\radius,0);
  \node at (8.5*\radius,0) {+};
  \draw[black, very thick] (9*\radius,0) -- (10*\radius,0);
  \draw[blue,fill=blue] (10*\radius,0,0) circle (.3ex);
  \draw[red, ultra thick] (10*\radius,0) -- (11*\radius,0);
  \draw[blue,fill=blue] (11*\radius,0,0) circle (.3ex);
  \draw[black, very thick] (11*\radius,0) -- (12*\radius,0);
  \node at (12.5*\radius,0) {+};
  \draw[black, very thick](13*\radius,0) --(14*\radius,0);
  \draw[blue,fill=blue](14*\radius,0,0) circle (.3ex);
  \draw[red,  thick](14*\radius,0) --(14.5*\radius,\radius);
  \draw[blue,fill=blue](14.5*\radius,\radius) circle (.3ex);
\draw[red, thick](14.5*\radius,\radius) --(15.5*\radius,\radius);
\draw[blue,fill=blue](15.5*\radius,\radius) circle (.3ex);
\draw[red, thick](14*\radius,0) -- (14.5*\radius,-\radius);\draw[blue,fill=blue](14.5*\radius,-\radius) circle (.3ex);
\draw[red,  thick](14.5*\radius,\radius) -- (14.5*\radius,-\radius);
\draw[red, thick](14.5*\radius,-\radius) -- (15.5*\radius,-\radius);
\draw[blue,fill=blue](15.5*\radius,-\radius) circle (.3ex);
\draw[blue,fill=blue](14.5*\radius,0) circle (.3ex);
\draw[red, thick](14.5*\radius,0) -- (15.5*\radius,0);
\draw[red,  thick](15.5*\radius,\radius) -- (15.5*\radius,-\radius);
\draw[blue,fill=blue](15.5*\radius,0) circle (.3ex);
\draw[red, thick](15.5*\radius,\radius) -- (16*\radius,0);
\draw[blue,fill=blue](16*\radius,0) circle (.3ex);
\draw[red, thick](16*\radius,0) -- (15.5*\radius,-\radius);
\draw[black, very thick](16*\radius,0) -- (17*\radius,0);
\node at (17.5*\radius,0) {+};
\draw[black, very thick](18*\radius,0) -- (19*\radius,0);
\draw[blue,fill=blue](19*\radius,0) circle (.3ex);
\draw[red, thick](19*\radius,0) -- (19.5*\radius,\radius);
\draw[red, thick](19*\radius,0) -- (19.5*\radius,-\radius);
\draw[blue,fill=blue](19.5*\radius,\radius) circle (.3ex);
 \draw[blue,fill=blue](19.5*\radius,-\radius) circle (.3ex); 
 \draw[red, thick](19.5*\radius,\radius) -- (20.5*\radius,\radius);
 \draw[red, thick](19.5*\radius,-\radius) -- (20.5*\radius,-\radius);
 \draw[blue,fill=blue](20.5*\radius,\radius) circle (.3ex);
 \draw[blue,fill=blue](20.5*\radius,-\radius) circle (.3ex);
 \draw[red, thick](20.5*\radius,\radius) -- (21*\radius,0);
 \draw[red, thick](20.5*\radius,-\radius) -- (21*\radius,0);
\draw[blue,fill=blue](21*\radius,0) circle (.3ex);
\draw[black,  thick](21*\radius,0) -- (22*\radius,0); 
\draw[red, thick](19.5*\radius,\radius) -- (20.5*\radius,-\radius);
\draw[red, thick](20.5*\radius,\radius) -- (19.5*\radius,-\radius);
\node at (22.5*\radius,0) {+};
\draw[black, thick](23*\radius,0) -- (24*\radius,0);
\draw[red, thick](24*\radius,0) -- (24.5*\radius,\radius);
\draw[red, thick](24*\radius,0) -- (24.5*\radius,-\radius);
\draw[blue,fill=blue](24*\radius,0) circle (.3ex);
\draw[blue,fill=blue](24.5*\radius,\radius) circle (.3ex);
\draw[blue,fill=blue](24.5*\radius,-\radius) circle (.3ex);
\draw[red, thick](24.5*\radius,\radius) -- (25.5*\radius,\radius);
\draw[red,  thick](24.5*\radius,-\radius) -- (25.5*\radius,-\radius);
\draw[red,  thick](25.5*\radius,\radius) -- (26*\radius,0);
\draw[red, thick](26*\radius,0) -- (25.5*\radius,-\radius);
\draw[blue,fill=blue](25.5*\radius,\radius) circle (.3ex);
\draw[blue,fill=blue](26*\radius,0) circle (.3ex);
\draw[blue,fill=blue](25.5*\radius,-\radius) circle (.3ex);
\draw[red,  thick](24.5*\radius,\radius) -- (24.5*\radius,-\radius);
\draw[red, thick](25*\radius,\radius) -- (25*\radius,-\radius);
\draw[red, thick](25.5*\radius,\radius) -- (25.5*\radius,-\radius);
\draw[blue,fill=blue](25*\radius,\radius) circle (.3ex);
\draw[blue,fill=blue](25*\radius,-\radius) circle (.3ex);
\draw[black, thick](26*\radius,0) -- (27*\radius,0); 
\node at (28*\radius,0) {+};
\node at (29*\radius,0) {...};

\end{tikzpicture}\right)\nonumber,\eea
\end{widetext}
which can be further recast in the following simplified form for the further analysis performed in this paper:
\newpage
\begin{widetext}
   \bea\Delta^{2}_{\zeta, {\bf RRR}}(k)
=A\times\left(1+\left(\frac{k}{k_s}\right)^2\right)\times\exp\bigg(6\ln\left(\frac{k_s}{k_e}\right)+\sum_{\textbf{All even graphs G}}{\cal F}_{\bf G}\bigg).\eea
\end{widetext}
Here the peak amplitude in the USR phase is given by:
\bea A&=&\bigg[\Delta^{2}_{\zeta,{\bf Tree}}(k_*)\bigg]_{\bf SRI}\times \left(\frac{k_e}{k_s}\right)^6.\eea
Here, the amplitude of the scalar power spectrum at the pivot scale is defined previously in equation (\ref{pivot}).
The primary outcome of the DRG resummed version of the one-loop corrected scalar power spectrum is that, unlike the previously derived renormalized one-loop power spectrum, it produces a controlled version of the two-point function after summing over all graphs contributing in even loop-order, where the logarithmic divergences' behavior is sufficiently softened to give a reliable computation. While the explicit details of the Feynman diagrams and the subgraphs are not necessary to perform the DRG resummation method, it is important to note that in the present context the leading order logarithmically divergent contributions originate from chain diagrams that continuously add cubic self-energy. The implementation of the DRG resummation does not need the dominance of all chain diagrams over other possible diagrams in the computation, but it will surely increase the leading contributions from these logarithmic-dependent components. It is rather amazing because without explicitly computing higher-loop corrections to the primordial power spectrum for the scalar modes, one may examine the behavior of each correction term in all-loop order. This implies that we may examine the spectrum's non-perturbative yet convergent behavior when the total all-loop contribution approaches a finite value, which in this case can be represented by an exponential function.

In figures (\ref{RRR1}) and (\ref{RRR2}), we have depicted the behaviour of the regularized-renormalized-resummed (RRR) version of the amplitude of the scalar power spectrum with respect to the number of e-foldings. In each of the plots, we have considered two possibilities of having EFT sound-speed parameters, $c_s=0.88$ and $c_s=1$ respectively, which strictly preserve the requirements of perturbativity, unitarity, and causality. Additionally, it is important to note that the RRR spectrum almost looks identical for the figures (\ref{RRR1}) and (\ref{RRR2}), which describe ekpyrotic ($\epsilon=7/2$) and matter ($\epsilon=3/2$) contraction-bounce scenarios, respectively. Comparing both of these plots we found that the behaviour in the ECP-EBP-SRI-USR-SRII in figure (\ref{RRR1}) is exactly same as found from the behaviour obtained in the MCP-MBP-SRI-USR-SRII in figure (\ref{RRR2}). It is further to point that, the total number of e-foldings covered in this analysis for both of these RRR version of the plots turns out to be, 
\begin{widetext}
   \bea \Delta N_{\bf Total}=\Delta N_{\bf C}+\Delta N_{\bf B}+\Delta N_{\bf SRI}+\Delta N_{\bf USR}+\Delta N_{\bf SRII}=25+10+16.2+2.3+7.5=61, \eea
\end{widetext}
which is a very important number to evade the {\it no-go theorem} proposed in \cite{Choudhury:2023vuj,Choudhury:2023jlt,Choudhury:2023rks} to generate solar or sub-solar mass PBHs in the present context of the discussion.
The prime reason for having identical features in both RRR versions of the figures (\ref{RRR1}) and (\ref{RRR2}) lies deeply in the quantum loop effects. It is important to note that, in both of these plots, loop corrections are highly suppressed in the contraction, bouncing, and in the SRI regions, as an immediate outcome of which the spectrum looks almost scale invariant in these regions after performing regularization-renormalization-resummation (RRR) procedures. On the other hand, a large deviation from the scale-invariant feature of the RRR spectra can be clearly visible in the USR and SRII phases, respectively. For each of the plots differences from the perspective of the EFT sound speed parameter can be also visible for $c_s=0.88$ and $c_s=1$ in the SRII region. 

\section{Primordial Black Hole Mass fraction from RRR Power Spectrum}\label{PBH}
In this section we review the formation of Primordial black holes as a result of the collapse of primordial density fluctuations with the equation of state parameter $(w)$. We will present a comparison of our model with the experimental data from EPTA and NANOGrav collaboration for different values of the EoS parameter $(w)$ and with different effective sound speeds $(c_s)$.
\begin{figure*}[htb!]\label{fpbhFig}
    \centering
    \subfigure[]{
    \includegraphics[height=8cm,width=8.5cm]{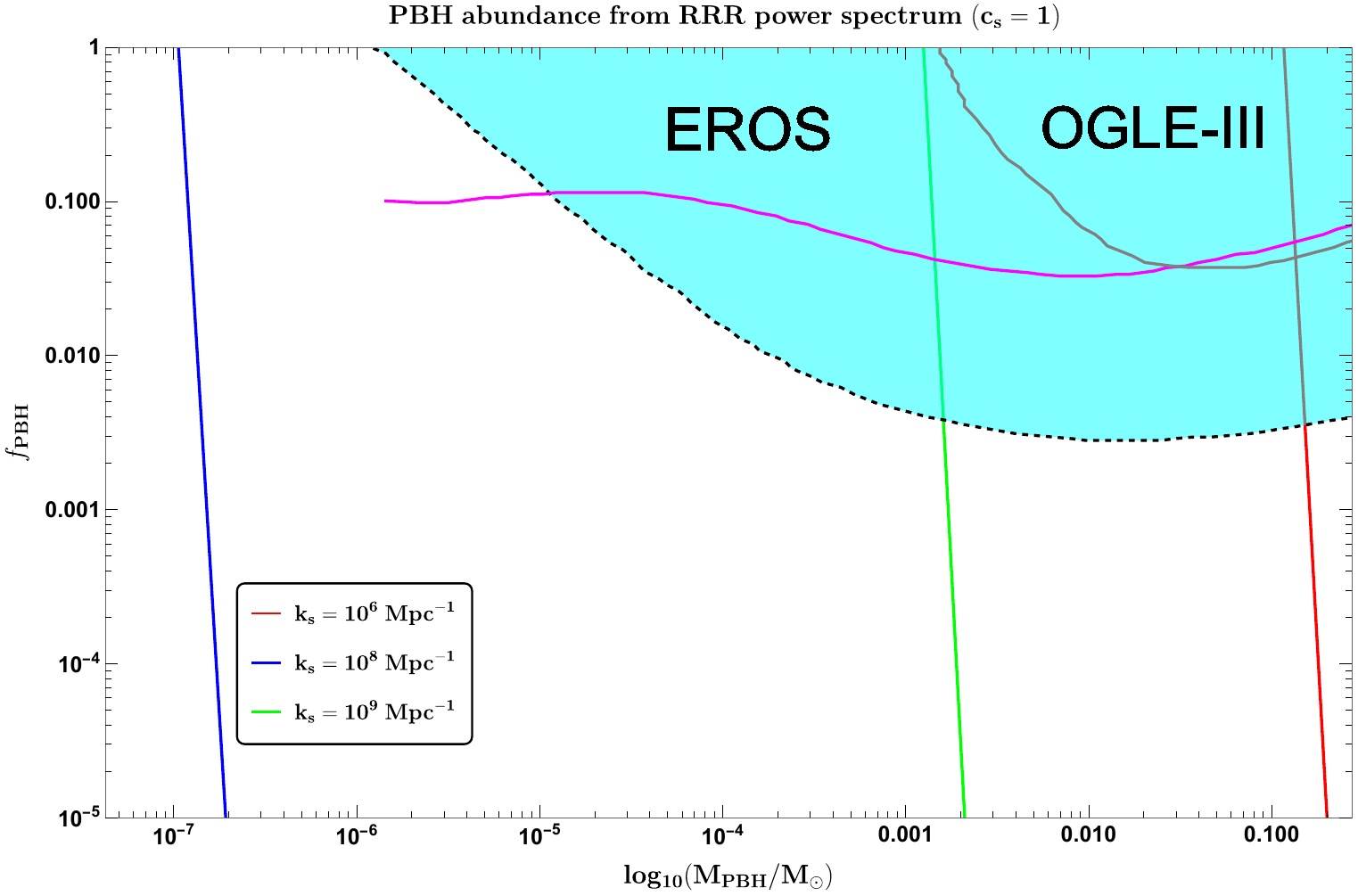}
    \label{fpbh_cs1}
    }
    \subfigure[]{
    \includegraphics[height=8cm,width=8.5cm]{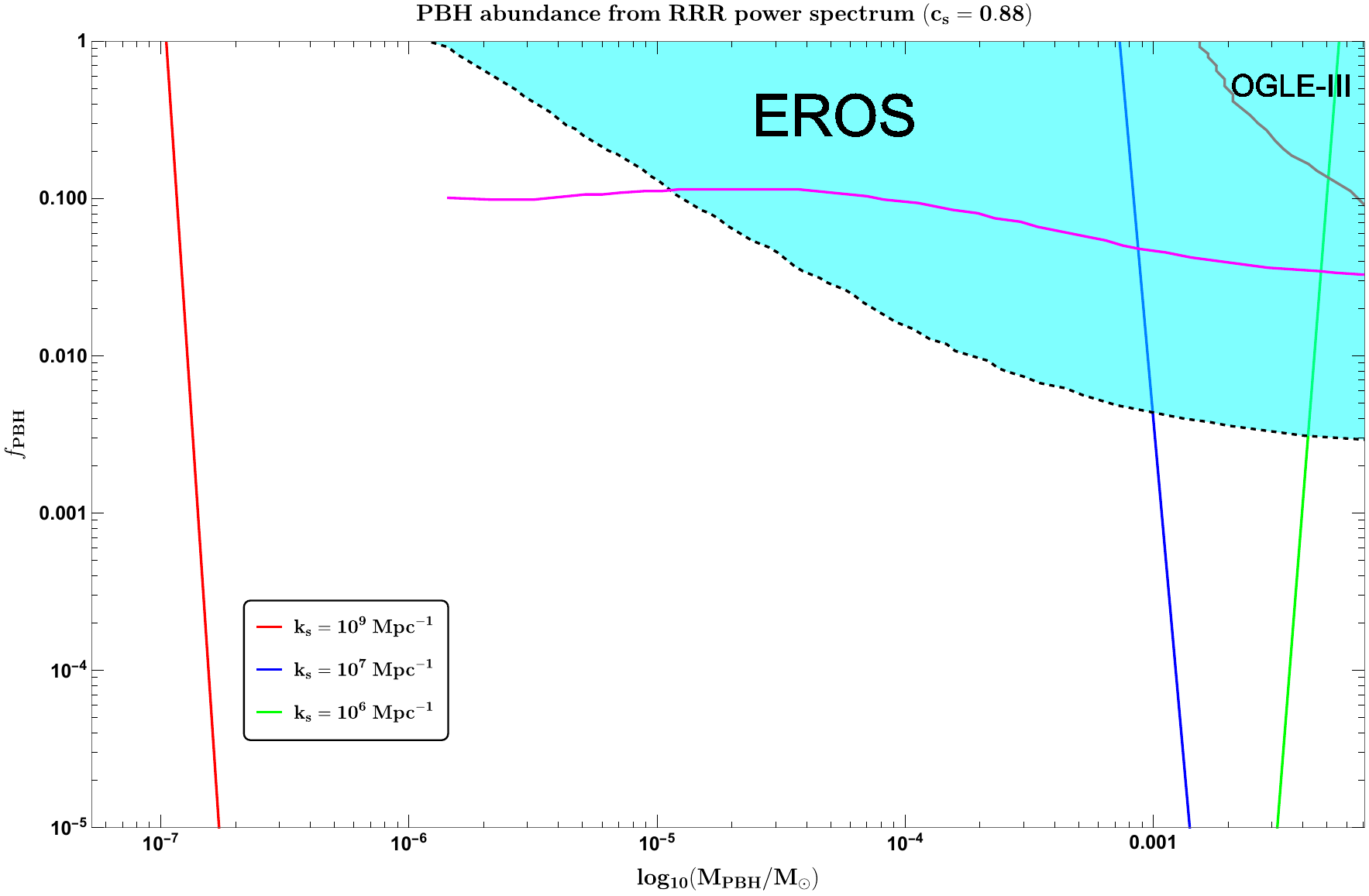}
    \label{fpbh_cs088}
    }
    \caption{Plots showing the fraction of PBH energy density, $f_{\rm PBH}$ as a function of their masses in solar mass units. The left one focuses on the mass limits when we fix $c_s=1$ and the right one focuses on the mass limits when we fix $c_s=0.88$. The different colour lines denote the different choice of transition wavenumbers. We have also included constraints coming from various microlensing experiments: cyan-coloured region highlights the recently obtained $95\%$ upper limits on PBH abundance with the dashed black boundary marking strict limits on $f_{\rm PBH}$ and also includes limits from other dark-matter surveys: EROS(magenta) and OGLE-III(gray).}
\end{figure*}

%\subsection{$w-$Press Schechter formalism}

This formalism deals with the formation of Primordial Black Holes when the perturbation over-density exceeds a certain threshold ($\delta \rho/\rho \equiv\delta >\delta_{th} $).\\
We will use the Carr's criterion of $c_s^2 =1 $ \cite{1975ApJ...201....1C} which gives us the relation of $w$ with the threshold as:
\bea\label{deltaTH}
\delta_{th} = \frac{3(1+w)}{5+3w},
\eea
The relation between the density contrast $(\delta(t,{\bf x}))$ with the comoving curvature perturbaion $\zeta$ assuming the linearity approximation between them in the super-horizon regime is:
\bea
\delta(k)\approx\frac{3(1+w)}{5+3w}\left(\frac{k}{aH}\right)^2\nabla^2\zeta(k),
\eea

The resulting mass of the PBH formed is \cite{Alabidi:2013lya}:
\bea\label{Mpbh}
M_{{\rm PBH}} = 1.13\time10^{15}\times\left(\frac{\gamma}{0.2}\right)\left(\frac{g_*}{106.75}\right)^{-1/6}\left(\frac{k_*}{k_s}\right)^{\frac{3(1+w)}{1+3w}}M_\odot,\nonumber
\eea
Where $\gamma\approx0.2$ is the efficiency of the collapse, $k_*=0.02{\rm Mpc}^{-1}$ is the pivot scale value and $M_\odot$ is the solar mass.
To estimate the PBH abundace, we need to find the variance of the distribution of the primordial overdensity, it is given by:

\bea\label{var}
\sigma_{\rm M_{PBH}}^2 = \left(\frac{2(1+w)}{5+3w}\right)^2\int\frac{dk}{k}(kR)^4W^2(kR)[\Delta_\zeta^2(k)]_{\rm \bf Total},\nonumber
\eea

Here $W(kR)$ is the Gaussian Smoothing function given by $e^{-k^2R^2/4}$, where $R=1/(c_sk_s)$.
We study our model with the following range of our parameters: $2/5\leq\delta_{th}\leq3$ and $-0.01\leq w\leq 1/3$. We will use these estimates to find their constraint which can produce a GW signal compatible with the NANOGrav15 data.
The mass fraction of the PBH is given by:
\bea
\beta(M_{\rm PBH}) =\gamma\bigg(\frac{\sigma_{M_{\rm PBH}}}{\sqrt{2\pi}\delta_{th}}\bigg){\rm exp}\bigg(\frac{-\delta_{th}^2}{2\sigma^2_{M_{\rm PBH}}}\bigg)
\eea
The PBH abundance is then given by:
\begin{widetext}
 \bea
    f_{\rm PBH} &=& \frac{\Omega_{\rm PBH}}{\Omega_{\rm CDM}}=1.68\times10^8\bigg(\frac{\gamma}{0.2}\bigg)^{1/2}\bigg(\frac{g_*}{106.75}\bigg)^{-1/4}\times(M_{\rm pbh})^{-\frac{6w}{3(1+w)}}\times\beta(M_{\rm PBH}),
    \eea   
\end{widetext}    
Here $g_*=106.75$ represents the relativistic degrees of freedom.
The frequency and wavenumber are related by the following relation:
\bea
f=1.6\times10^{-15}\bigg(\frac{k}{{\rm Mpc^{-1}}}\bigg),
\eea
Following are the graphs of PBH formation and their analysis.
%\subsection{Numerical Outcomes I: Constraints of PBH Formation}

This section also focuses on the numerical outcomes of PBH formation. They will be confronted with the recent observational constraints coming out of the detailed microlensing experiments data analysis. We will use regularized-renormalized-resummed scalar power spectrum to perform our analysis. In fig.\ref{fpbh_cs1} and fig.\ref{fpbh_cs088} we look for the behaviour of $f_{\rm PBH}$ as their mass changes for different values of $k_s$ and $c_s$. For both the figures (left and right) the value of $f_{\rm PBH}$ changes very quickly to $f_{\rm PBH}\lesssim10^{-4}$ as the value of mass changes for both $c_s=1$ and $c_s=0.88$. We see that $f_{\rm PBH}$ becomes neglegible  with $M_{\rm PBH}$ in a very short interval where the different masses are considered on the basis of the choice of different wavesnumbers $k_s$.

 \section{Scalar Induced Gravitational Waves with general equation of state}\label{SIGW}

This section aims to review the theory of gravitational waves produced by scalar fluctuations. In the cosmic perturbation theory, second-order induced GWs are produced by mode couplings of the first-order perturbations to the FLRW metric. Compared to the CMB fluctuations, this effect creates a considerable amplification in the measured GW spectrum, indicating the presence of a massive scalar disturbance. We are investigating the case when a state with an unknown equation of state and a value of $w$ coexists with the end of inflation. Deep within the horizon in this broad $w$ backdrop, where the modes are mostly sourced, the induced GW formed in the early cosmos can exist. Taking that into consideration, the induced GW spectrum for a generic $w$ backdrop is expressed therefore:
\bea
\label{GWden}
\Omega_{\rm{GW},0}h^2 &=& 1.62 \times 10^{-5}\;\bigg(\frac{\Omega_{r,0}h^2}{4.18 \times 10^{-5}}\bigg)\nonumber\\
&&\times\bigg(\frac{g_{*}(T_c)}{106.75}\bigg)\bigg(\frac{g_{*,s}(T_c)}{106.75}\bigg)^{-4/3}\Omega_{\rm GW},
\eea
 Here, the radiation energy density as it is measured today is denoted by $\Omega_{r,0}h^2$, and the effective degrees of freedom for energy and entropy are $g_{*},g_{*,s}$. When such generated GWs behave as freely propagating GWs throughout the EoS $w$-dominated period, the number $\Omega_{\rm GW}$ reflects the GW energy density fraction.

Using the kernel functions for the modes and scales that fulfill $k \geq k_{*}$, we further describe the energy density $\Omega_{\rm GW}$ as follows \cite{Domenech:2021ztg}:
\begin{widetext}
\bea
\label{omegaGW}
\Omega_{\rm GW} &=& \frac{k^{2}}{12a^{2}H^{2}}\times\overline{\Delta^{2}_{h}(k,\tau)} = \left(\frac{k}{k_*}\right)^{-2b}\int_0^\infty dv \int_{|1-v|}^{|1+v|} du \; {\cal T}(u,v,b,c_s)\times\overline{\overline{\Delta^2_{\zeta,\bf{EFT}}(ku) }} \times\overline{\overline{\Delta^2_{\zeta,\bf{EFT}}(kv)}},\label{eq:1}
\eea
\end{widetext}

Where $b$ is an EoS-dependent parameter defined to be as $b=(1-3w)/(1+3w)$ and $k_*$ is the usual pivot scale (CMB).

The transfer function in the Eqn.(\ref{omegaGW}) for the general case of constant values of the EoS $w$ and the propagation speed $c_s$ is given by:
\begin{widetext}
 \bea\label{transferFunc}
    {\cal T}(u,v,b,c_s)&=& {\cal N}(b,c_s)\left[ \frac{4v^2-(1-u^2+v^2)^2}{4u^2v^2}\right]^2\times|1-y^2|^b\times\bigg\{(P_b^b(y)+\frac{b+2}{b+1}P_{b+2}^{-b}(y))^2\Theta(c_s(u+v)-1)\nonumber\\&&+\frac{4}{\pi^2}\left[
    (Q_b^{-b}(y)+\frac{b+2}{b+1}Q_{b+1}^{-b}(y))^2\Theta(c_s(u+v)-1) +(Q_b^{-b}(y)+2\frac{b+2}{b+1}Q_{b+1}^{-b}(-y)\Theta(1-c_s(u+v))) \right]
    \bigg\}.\quad\quad
\eea   
\end{widetext}
A propagation speed dependant parameter $y$ has been introduced for simplicity and given as:
\bea\label{y}
y\equiv1-\frac{1-c_s^2(u-v)^2}{2c_s^2uv},
\eea
The expression also contains a normalized EoS dependant parameter given as:
\bea
{\cal N}(b,c_s)\equiv\frac{4^{2b}}{3c_s^4}\Gamma\left(b+\frac{3}{2}\right)^4\left(\frac{b+2}{2b+3}\right)^2(1+b)^{-2(1+b)},\nonumber
\eea
Further details are presented in Appendix(\ref{A1}).

\section{PBH Overproduction and its possible resolution}\label{OP}
We have attempted to address and identify the energy density spectrum of induced GW production in a broad EoS $w$ context in the preceding section. A PBH counterpart can be connected to the large boosts resulting from scalar mode couplings in the very early Universe. The observation of SIGW may also indicate the possibility of producing near solar-mass PBH in the very early Universe due to the elevated power spectrum that is sensitive to the scales of NANOGrav15 and EPTA.

Regarding the issue of PBH overproduction, there has been a lot of progress recently. Many people investigated the SGWB signal that the PTA partnerships published to determine whether it had astronomical or cosmic origins. The SIGW model is one of the many potential outcomes that most closely matches the PTA data, yet it was shortly proposed that this scenario could have an overproduction issue. Many people have actively worked to find an explanation and remedies for this problem; see \cite{Ferrante:2023bgz,Franciolini:2023pbf,Gow:2023zzp,Gorji:2023sil,Firouzjahi:2023xke} for relevant material. The main issue here is that attaining a significant abundance of near-solar-mass PBH, which coincides with the frequencies investigated by the NANOGrav15 signal, ultimately results in a conflict with the SIGW interpretation of the same SGWB signal, which is supported by the PTA partnerships. Because the augmented amplitude becomes $\sim {\cal O}(1)$ during the domain associated with PBH generation, forcing higher statistical agreement with PTA will inevitably result in a collapse of perturbation theory. From the standpoint of closely matching the data, the aforementioned makes the problem more problematic. Either a different interpretation of the signal itself or one that can more closely resemble the SIGW interpretation of the data is required in order to avoid overproduction. The authors in \cite{Franciolini:2023pbf,Ferrante:2022mui} add non-linearities to the density contrast that exists in the super-Horizon regime and consider the effect of non-Gaussianities to bolster their assertions concerning overproduction. The problem becomes more inflexible and requires more attention as a result of these theoretically required elements and a thorough examination across several models. Higher-order estimations of abundance in the super-Horizon and the appropriate density contrast threshold range are two more essential concerns concerning the precise theoretical calculations for the PBH abundance. It is anticipated that these will not materially change the current study and its findings, but they still need to be firmly proven.  

As previously stated, there has been a current push by several writers to offer appealing remedies to the overproduction problem for PBHs. The curvature perturbations at PBH-relevant scales are produced by a potential curvaton scenario, which might have existed as a spectator field in the very early Universe. It's also possible that the detected signal was caused by an additional tensor spectator field. Due to the participation of non-Gaussianities, which can become extremely sensitive to the scalar power spectrum amplitude, the presence of non-attractor characteristics in the theory of concern is also essential for having abundant PBH generation. In this part, we focus on the non-Gaussian aspects of the primordial density perturbations, which are crucial in understanding PBH generation. Furthermore, given the overproduction problem, the rationale behind selecting this scenario will be covered in this part, as we are working with an arbitrary EoS value. 

A common scenario is including an ultra-slow roll (USR) phase into inflation or its alternative bouncing  frameworks. This introduces significant quantum fluctuations that produce primordial curvature perturbations, which eventually collapse into PBHs or, in some situations, create gravitational waves. Although we assume a Gaussian profile for the power spectrum of  curvature perturbations for the sake of simplicity, including non-Gaussianity enables a thorough and in-depth examination. For example, the probability distribution function (PDF) \cite{Kawaguchi:2023mgk,DeLuca:2022rfz,Taoso:2021uvl,Atal:2018neu,Young:2013oia,Byrnes:2012yx,Bullock:1996at}, which is used to calculate the PBHs abundance, is significantly affected by non-Gaussianity. A comprehensive examination of the PDFs associated with single-field inflation, as presented in Ref.\cite{Pi:2022ysn}, indicates the existence of a logarithmic relation for e-fold fluctuations, $\delta N$, when a non-attractor regime is present during inflation. The predominance of these logarithms in $\delta N$ can cause the PDF to rapidly develop exponential tails, which have the potential to substantially impact the PBH mass fraction. Using the threshold statistics of the compaction function in linear cosmological perturbation theory, we addressed the non-linearity and non-Gaussianity related to the curvature perturbations in ref.\cite{Choudhury:2023fwk}, working within the framework of Galileon theory. We took into account the significant negative local non-Gaussianity connected to the initial disturbances. According to our research, PBH masses up to $M_{\rm PBH} \sim {\cal O}(10^{-3}-10^{-2} M_{\odot})$, taking into account $f_{\rm NL} \sim -6$, might prevent PBH overproduction.

\begin{figure*}[htb!]
    \subfigure[]{
    \includegraphics[height=8cm,width=8.5cm]{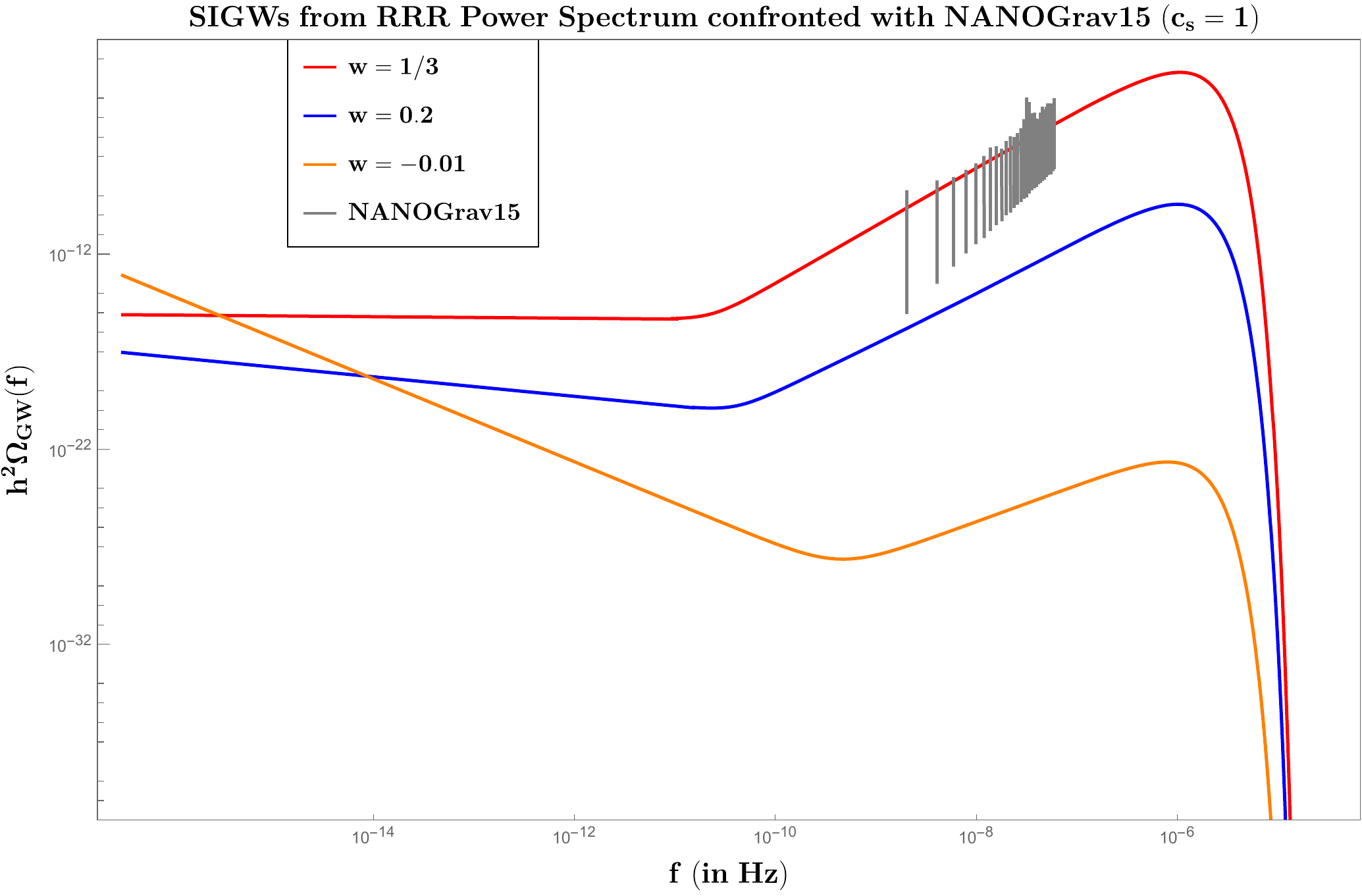}\label{wNG1}
    }
    \subfigure[]{
    \includegraphics[height=8cm,width=8.5cm]{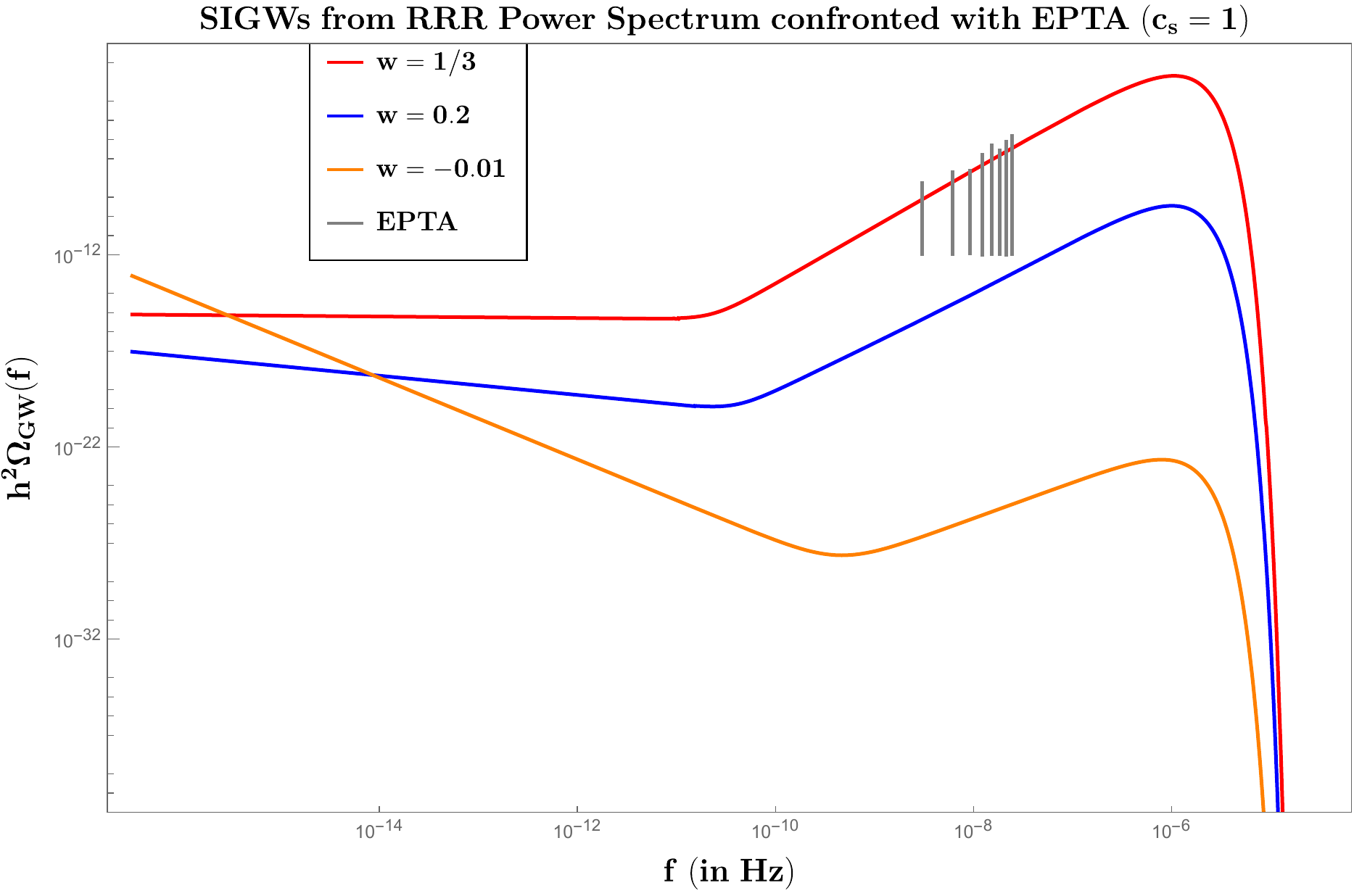}\label{wEPTA1}
    }
    \subfigure[]{
    \includegraphics[height=8cm,width=8.5cm]{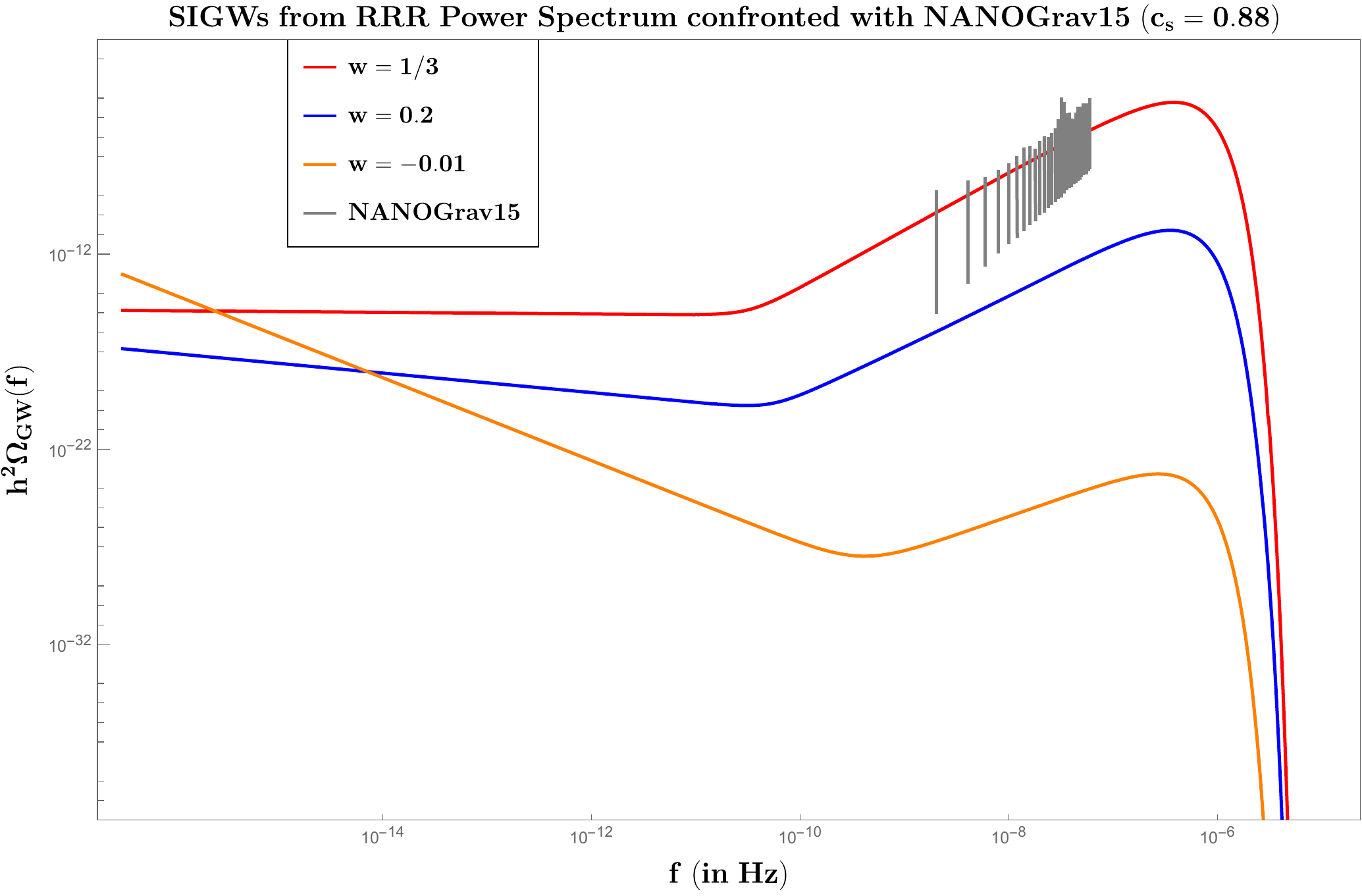}
    \label{wNG088}}
    \subfigure[]{
    \includegraphics[height=8cm,width=8.5cm]{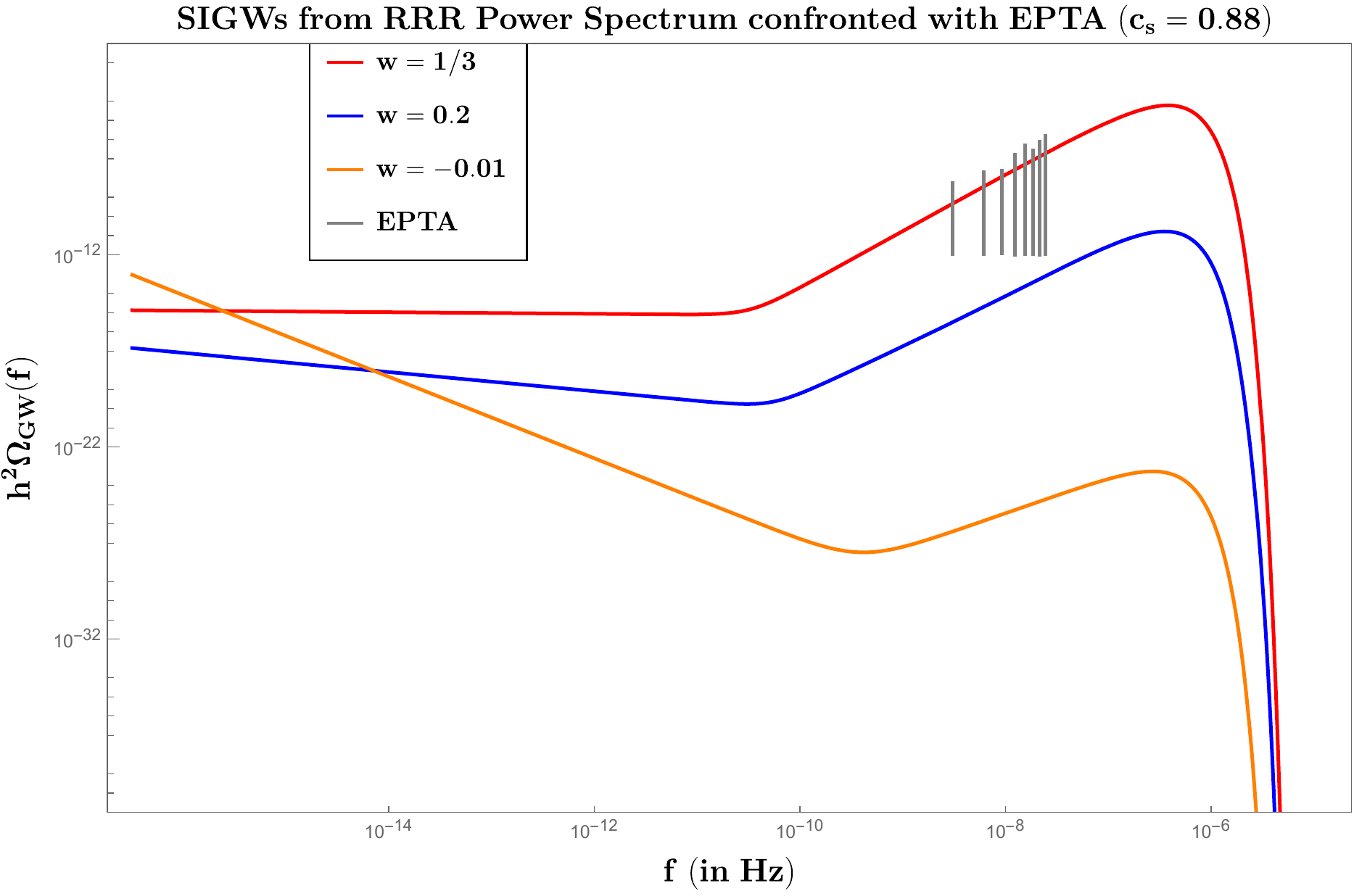}
    \label{wEPTA088}}
    \caption{$h^2\Omega_{GW}$ for various EoS parameters ($w$) plotted as a function of frequency and confronted with observational data from NANOGrav15 and EPTA for different values of the propagation speed ($c_s$).}\label{wSIGW}
\end{figure*}

\begin{figure*}[htb!]
    \centering
    \subfigure[]{
    \includegraphics[height=8cm,width=8.5cm]{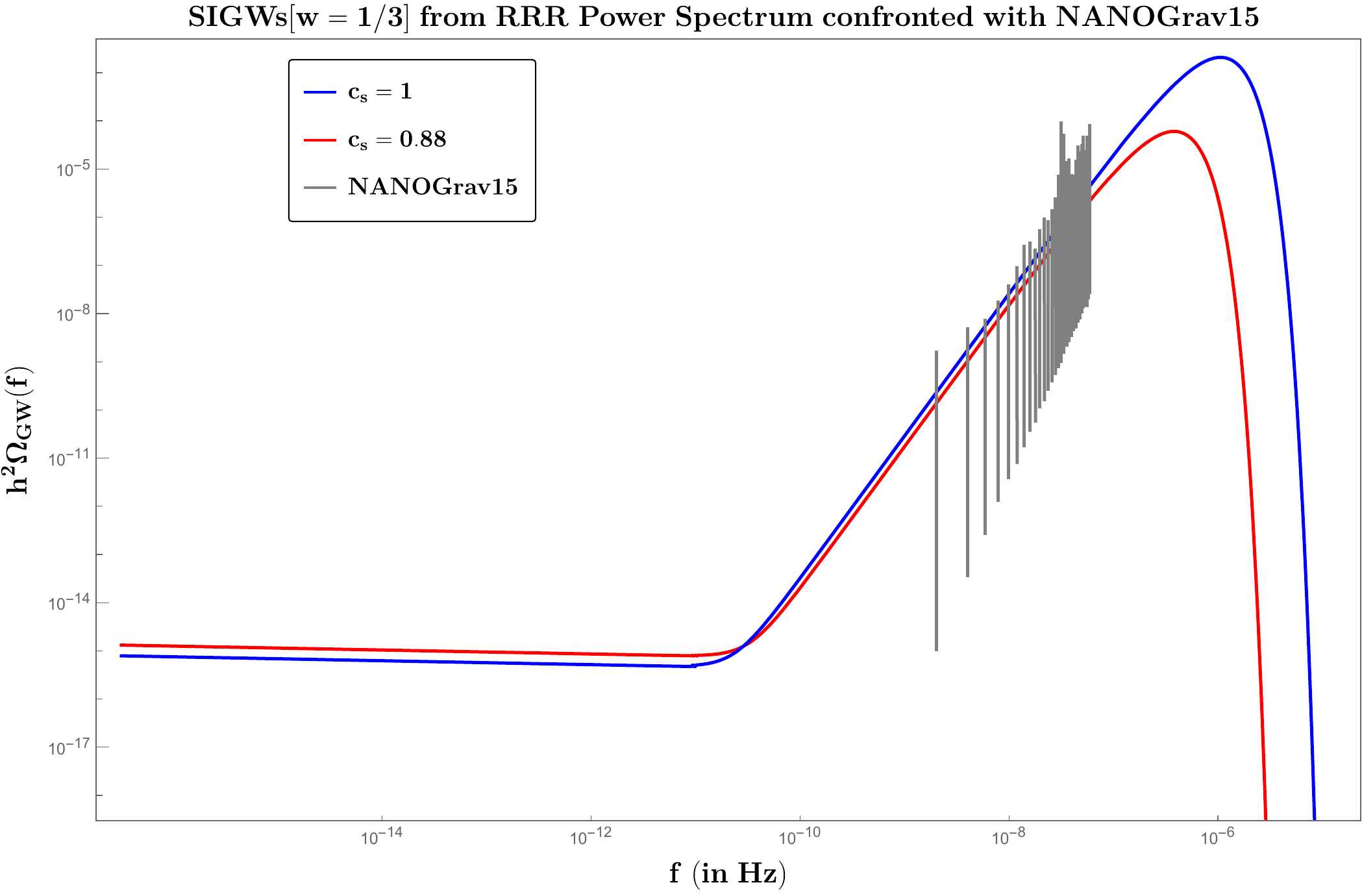}\label{DiffcsNG}
    }
    \subfigure[]{
    \includegraphics[height=8cm,width=8.5cm]{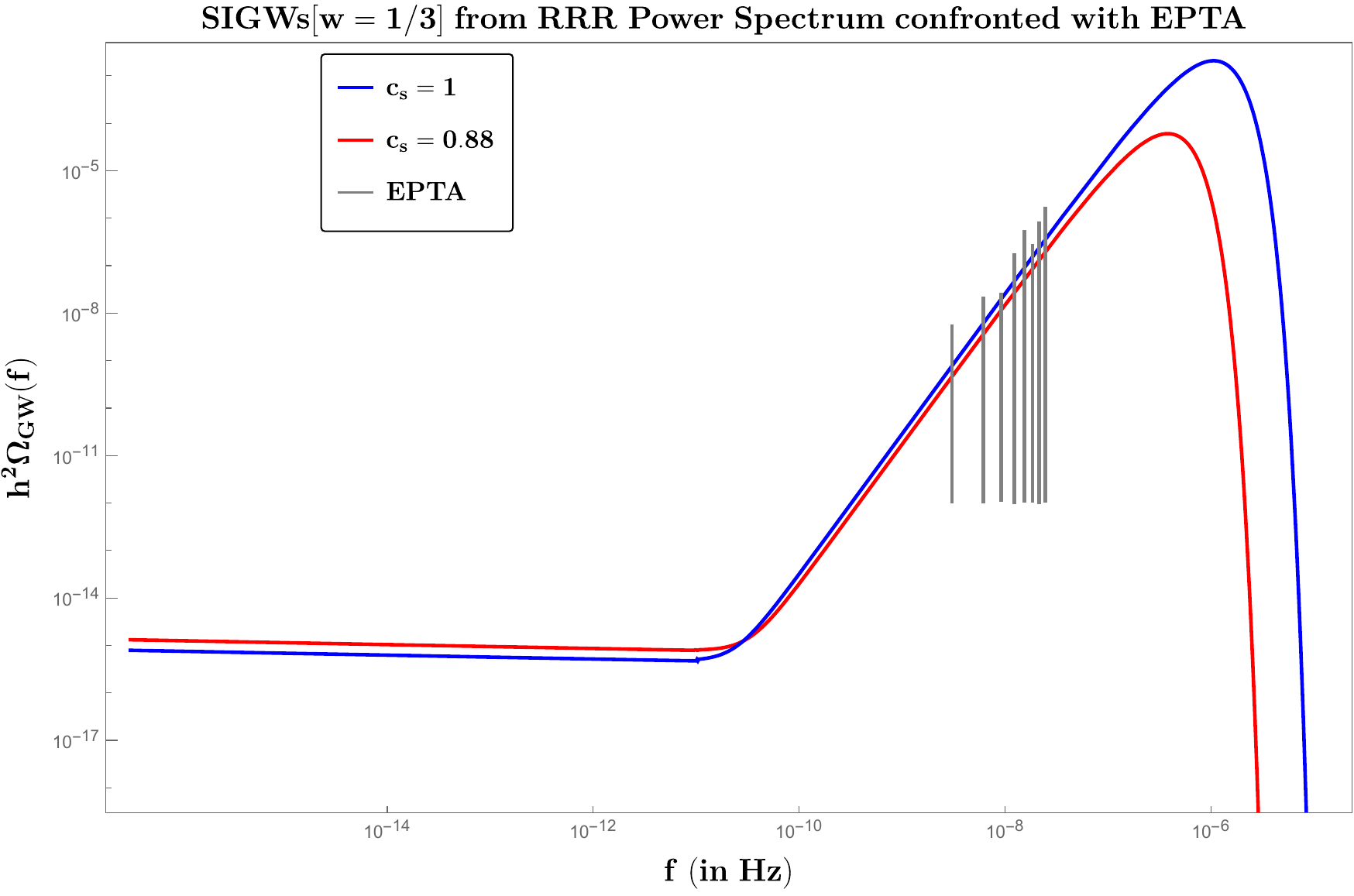}\label{DiffcsEPTA}
    }
    \caption{$h^2\Omega_{\rm GW}$ plotted as a function of frequency for different values of the propagation speed $s_s$ keeping the EoS parameter fixed $w=1/3$. We have also included the experimental data from NANOGrav15 (left) and EPTA (right).}\label{csSIGW}
\end{figure*}
    \begin{figure*}[htb!]
    \centering
    \subfigure[]{
    \includegraphics[height=8cm,width=8.5cm]{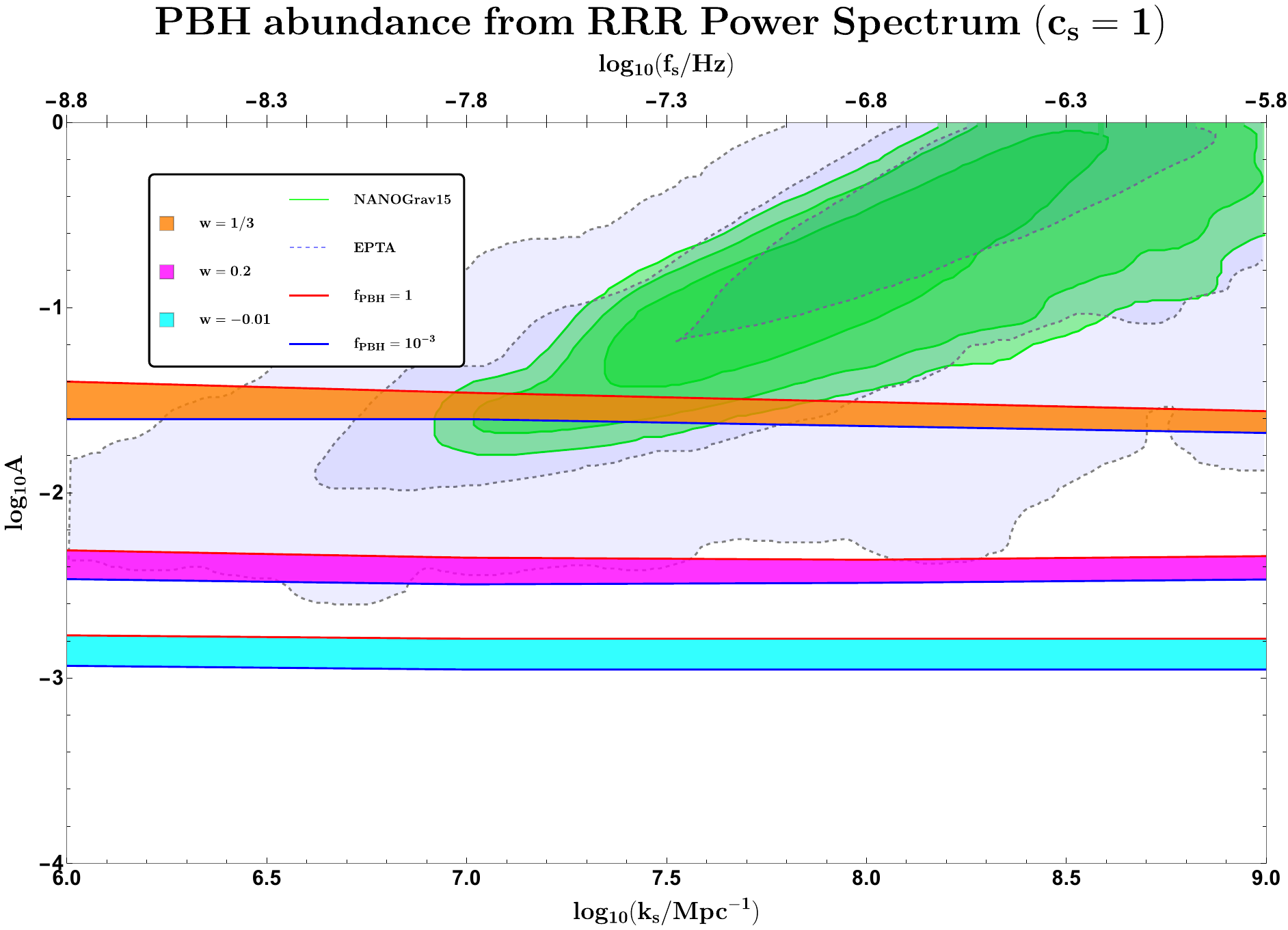}\label{over1}
    }
   \subfigure[]{
   \includegraphics[height=8cm,width=8.5cm]{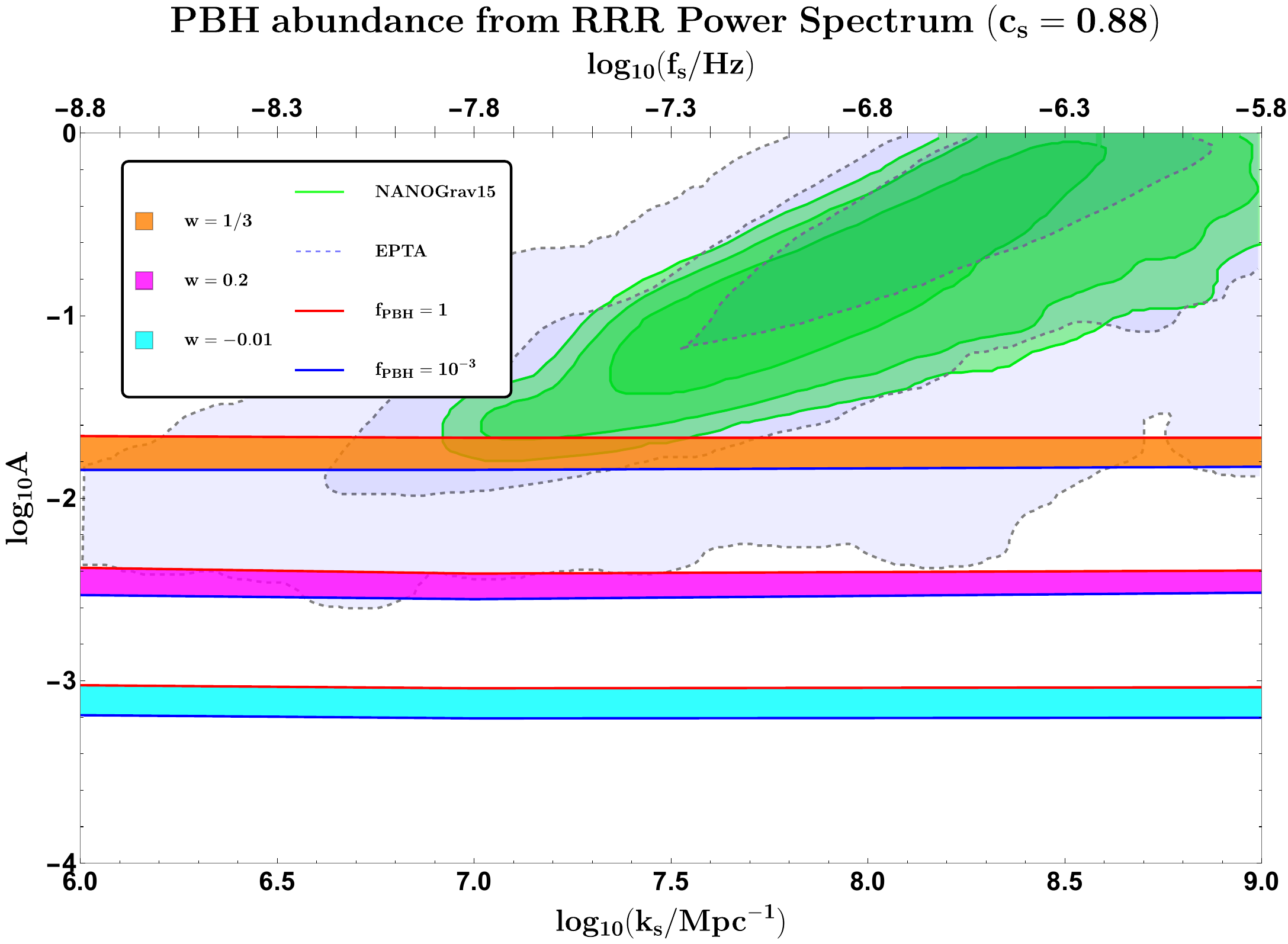}
\label{over088}
   }
   \caption{The above figure shows the change in the peak amplitude of the RRR scalar power spectrum with changing transition wave number for two different values of $c_s$. The bands enclose the region corresponding to the abundance $f_{\rm PBH} \in (1,10^{-3})$. The posteriors for the NANOGrav15 and EPTA data are represented in green and light blue respectively. }\label{OPA}
\end{figure*}
This offers an interesting field of study, allowing us to explore different constant EoS parameter combinations and explore potential PBH formation situations. Within the linear domain of the perturbation theory, we have studied instances when the EoS parameter $w$ has discrete values, that is, $w\in \{-0.01, 0.2, 1/3\}$. We also confine our considerations to the cases in which we maintain $c_{s}^{2}=1$ for the propagation speed. Considering the well-known PBH generation mechanism derived from critical collapse theory, the case of an RD period has been selected up to this point. Nonetheless, we take use of the freedom that comes with the lack of knowledge about the values for EoS parameter in the very early Universe to study the potential impacts of the huge fluctuations that enter such a period on the induced GWs and PBH production that follows. We obtain a more robust image of the cosmic backdrop where various cosmological events can occur by comparing the outcomes of our study with the new data. We may strengthen the case for $w=1/3$ relative to the consequences of alternative values of $w$ by utilizing our work with EFT of non-singular bounce. This research has a direct impact on the mass fraction and SIGW spectrum of the near solar-mass PBH. We discover that the overproduction avoiding situations are within $1\sigma-2\sigma$ when faced with NANOGrav15, and the EoS parameter $w$ lies in the range $(0.31, 1/3)$  such that we obtain $f_{\rm PBH} \in (10^{-3},1)$. For PBH masses that match the sizes of the NANOGrav15 signal, we do not acquire adequate abundance for $w$ values below the range mentioned above. For more details see the figure \ref{OPA} and related discussions in the next section.

Therefore, in practice, we deal with changes in two distinct threshold variables: (a) the density contrast threshold when examining the super-Hubble scales linear regime approximation, in which non-Gaussianities are absent (this work and some other earlier work \cite{Choudhury:2024one,Choudhury:2023fjs,Bhattacharya:2023ysp,Choudhury:2023hfm}); and (b) the compaction threshold when taking into account the non-linearities inherent in the super-Hubble and the non-Gaussianities, $f_{\rm NL}$, which arise as a part of this in the comoving curvature perturbation (earlier work, see ref \cite{Choudhury:2023fwk}). 

We stress once more that our approach is valid only when considering the density contrast and operating within the linear domain. Under this supposition, the Press-Schechter formalism adjusted for the EoS $w$ indicates that a value of $w$ in the range (0.31,0.33) is the most likely solution to the overproduction problem. Incorporating non-Gaussianities, $f_{\rm NL}$, and non-linear statistics of the density contrast into the analysis of the PBH mass fraction strengthens the present analysis with the EoS. Slow-roll violation causes the curvature perturbations during PBH generation to generally defy Gaussian statistics; to account for this, a local perturbative expansion of the following kind is often selected:
\bea
\zeta({\bf x}) = \zeta_{G}({\bf x}) + \frac{3}{5}f_{\rm NL}(\zeta_{G}^{2}({\bf x}) - \langle\zeta_{G}^{2}({\bf x})\rangle) + \cdots
\eea 
Such a non-Gaussianity factor can have a significant effect on PBH creation.
To wrap off our conversation, we want to draw attention to the specific non-linear relationship that was previously discussed:
\begin{widetext}
    \bea \label{NL}
\delta({\bf x},t) &=&-\frac{2(1+w)}{5+3w}\left(\frac{1}{aH}\right)^{2}e^{-2\zeta({\bf x})}\bigg[\nabla^{2}\zeta({\bf x}) + \frac{1}{2}\partial_{i}\zeta({\bf x})\partial_{i}\zeta({\bf x})\bigg],
\eea
\end{widetext}
See reference \cite{Young:2024jsu} for more details. The non-Gaussian statistics for $\delta({\bf x},t)$ must be taken into account, and when combined with the local non-Gaussianity, $f_{\rm NL}$, can drastically change the analysis for the PBH mass fraction. See refs. \cite{Choudhury:2023fwk, Young:2019yug, DeLuca:2019qsy} for more details.

%Primordial Black Holes are formed due to the gravitational collapse of the curvature perturbations upon horizon re-entry. If this curvature perturbation is too large, then this can cause the PBH abundance to exceed that of the Dark Matter, implying that all the dark matter in our universe is made up of PBHs, which is against the observed data. Therefore, we tweak our model parameters to circumvent the overproduction problem.\\
%In this paper, we vary the transition wavenumber $k_s$ which lies inside the NANOGrav15 frequency spectrum for different EoS parameter($w$). From this analysis we determine which $w$ is most favourable to solve the overproduction problem. We perform this analysis for different values of $c_s$ (1 and 0.88).

\section{Numerical Outcomes and Discussions}\label{Num}

In this section, we will address the overproduction problem. 
PBHs are formed by the collapse of curvature perturbations as they re-enter the horizon. If the perturbations are too large, it could lead to the abundance of PBH exceeding that of present dark matter abundance. This raises a concern as PBHs are a candidate for dark matter. The abundance of PBHs being greater than that of dark matter would imply that the dark matter that we observe today is made entirely of PBHs which goes against the observational imprints. Therefore we need to tune our model in such a way that this overproduction issue is dealt with. One way to address this issue is through the lens of the EoS parameter, which we do here.

The agenda of this paper is to find out which value of $w$ will be more favorable to solve the overproduction problem. We will consider the effects of the EoS parameter $w$ on the PBH abundance, which are formed in the region of frequencies that are sensitive to NANOGrav15 signal. We vary the transition wavenumber $k_s$ within the range of values coinciding with NANOGrav15 frequency spectrum, and then compute the corresponding peak amplitude of the scalar power spectrum which gives us the preferable PBH abundance for different values of $w$. We plot the results along with the data curves from NANOGrav and EPTA. We perform the above analysis for different values of the propagation speed $c_s$. Finally, we discuss the results of the SIGW spectrum for multiple values of $w$ with the data from NANOGrav and EPTA.

The plots in fig.(\ref{wSIGW}) show the SIGW spectrum for various values of the EoS parameter $w$ plotted along with data from NANOGrav and EPTA. The plots are carried out while keeping the values of the propagation speed at $c_s=1$ and $c_s=0.88$.
We observe that as we decrease the value of $w$, the peak amplitude of the spectrum of SIGW falls off quickly. In each case, the value $w=1/3$ has the highest value of peak amplitude and it falls under NANOGrav and EPTA region. The spectrum observed for $w=0.2$ \;and $w=-0.01$ lies outside NANOGrav and EPTA region. The values of $w$ close to 1/3 are the most favourable ones.

The plots in fig.(\ref{csSIGW}) show the change in $\Omega_{\rm GW}$ when we change the value of the propagation speed $c_s$ while keeping $w$ constant at 1/3. Decreasing the value of $c_s$ makes the peak amplitude of the spectrum fall off but it still lies inside the region from the data given by NANOGrav and EPTA. So, changing the value of $0.88<c_s<1$ does not have any significant effect on the region of the spectrum sensitive to NANOGrav15 and EPTA. The value of $c_s$ cannot exceed $1$ as it would break unitarity, perturbativity, and causality. The lower bound for $c_s$ is found to be $0.024$ by Planck data \cite{Planck:2018jri}. The value of $\Omega_{\rm GW}$ in fig.(\ref{csSIGW}) does not change very much during the contraction, bounce and SRI phase. It undergoes an increase in value during the USR phase and falls of rapidly during SRII phase due to the negative exponential nature of the RRR power spectrum.

Fig.(\ref{OPA}) shows how the peak amplitude $A$ of the RRR scalar power spectrum behaves with increasing the transition wavenumber sensitive to PTA experiments. The value of $c_s$ is set to $1$ in fig.(\ref{over1}) and to $0.88$ in fig.(\ref{over088}). We have chosen a set of benchmark values for the EoS parameter, $w\in\{-0.01,0.2,1/3\}$. This interval leads to the allowed density contrast to be in the range $0.59\leq\delta_{th}\leq2/3$ and the interval for EOS to be $-0.01\leq w\leq 1/3$ from eq.(\ref{deltaTH}). If we look into fig.(\ref{over1}), the case of $w=1/3$ is highly preferred for the effective sound speed parameter $c_s=1$, and the corresponding orange band lies in the $2\sigma$ region of the NANOGrav15 signal. As we move lower in the value of $w$, to $w=0.2$ and $w=-0.01$, the magenta and cyan bands corresponding to the respective values lie quite outside the $3\sigma$ region. In fig.(\ref{over088}), where we have set $c_s=0.88$, we see that the orange band corresponding to $w=1/3$ lies inside the $3\sigma$ region of the NANOGrav15 signal and both the bands for $w=0.2$ and $w=-0.01$ lies quite outside the NANOGrav15 signal. The bands have shifted down after lowering the value of $c_s$ from 1 to 0.88 because the peak amplitude of the scalar power spectrum decreases by decreasing the value of $c_s$. We see a concurring trend where the band corresponding to higher value of $w$ is placed higher in the graph. We can keep increasing the value for $w$ to get the band in the $1\sigma$ region but by doing this, the spectrum for SIGW does not match the NANOGrav and EPTA data. In the case where $w>1/3$, the corresponding  peak amplitude of the scalar power spectrum turns out to be greater than unity, which would break the underlying predictions of cosmological perturbation theory. Therefore, increasing the value of $w$ would lead to inconsistencies in the present analysis. In both cases ($c_s=0.88,1$), we see that the value of $w$ lies in a tiny constrained window, $0.31<w<1/3$. Similarly decreasing the value of $w$ below 0.31, the amplitude of the scalar power spectrum would go below a value that lies outside the $3\sigma$ contour of the NANOGrav15 signal. 

\section{Conclusion}\label{con}

This paper discusses the uncertainty regarding the precise nature of the equation of state parameter (EoS) $w$ immediately prior to the beginning of the BBN and its implications for the severe PBH overproduction problem when compared to the observed GW energy density spectrum from the PTA collaboration (which includes both NANOGrav15 and EPTA). This theoretical paradigm is based on an EFT of non-singular bounce. After combining the regularization, renormalization, and resummation procedures, the model is utilized to analyze the formation of PBHs and the power spectrum of the scalar modes, where the essential one-loop contributions are taken care of. The findings from the analysis of the impacts of changing effective sound speed values provide a crucial piece of information that is used in all of the talks. The critical interval of $0.88 \leq c_{s} \leq 1$, which aids in achieving the best characteristics in the DRG-resummed and one-loop renormalized scalar power spectrum suitable for studying PBH generation, is provided by this analysis. We performed the analysis of GW spectrum research for arbitrary but constant $w$, considering the case where the density contrast fluctuations are dominated by linearities. In these circumstances, we discovered that the current permissible interval on the density contrast threshold from numerical research is $2/5 \leq \delta_{\rm th} \leq 2/3$. From our EoS interval $-0.01 \leq w \leq 1/3$, we found that the range for our density contrast threshold lies in the interval $(0.59, 2/3)$ which is a subset of the permissible interval. Restricting the following analysis of this work to this range for $w$, we observed that the situation of $0.31\lesssim w \leq 1/3$ is the most favorable to get a significant PBH abundance of $f_{\rm PBH} \in (1,10^{-3})$ with large mass PBHs, $M_{\rm PBH}\sim {\cal O}(10^{-7}-10^{-3})M_{\odot}$, after imposing the constraint of no overproduction. When confronted with PTA, we find that the overproduction avoiding circumstances are between $1\sigma-2\sigma$, while the EoS parameter lies inside the narrow window, $0.31<w\leq 1/3$. The resultant amplitude is severely suppressed for the values of $w<0.2$, which reduces its usefulness in explaining any observationally meaningful data as originating from the investigation of EFT of non-singular bounce. We emphasize once more the need for using linear approximations in our analysis when utilizing the density contrast $(\delta \equiv \delta\rho/\rho)$ to calculate the PBH abundance and SIGW spectra. We also point out that correct PBH abundance analysis is still lacking when considering an arbitrary EoS $w$ because one must account for the non-linearities in $\delta$ on the superhorizon scales and the non-Gaussianities that are subsequently developed in the corresponding distribution function. We intend to deal with this construction shortly. Another potential subject for future research, on which we want to develop, is what modifications could take place in the current analysis of arbitrary constant $w$ when a smooth transition into the USR phase replaces a sharp transition. Finally, but just as importantly, we want to include non-linearities and non-Gaussianities in the framework of the compaction function in the presence of the now mentioned technique with broad EoS.

\section*{Acknowledgement}

SC would like to thank The North American Nanohertz Observatory for Gravitational Waves (NANOGrav) collaboration and the National Academy of Sciences (NASI), Prayagraj, India, for being elected as an associate member and the member of the academy respectively. SC would like to acknowledge the inputs from The North American Nanohertz Observatory for Gravitational Waves (NANOGrav) collaboration members, for useful comments and discussions, which helped the improvements of the presentation of the article. Also, SC would like to especially thank Soumitra SenGupta and Supratik Pal for inviting to IACS, Kolkata, and ISI, Kolkata, during the work. Furthermore, SC thanks Supratik Pal and his students for inviting them to give an inaugural plenary talk at the discussion meeting titled {\it Cosmo Mingle}, where part of the work was presented. SC would also like to thank all the members of Quantum Aspects of the Space-Time \& Matter
(QASTM) for elaborative discussions. SP is supported by the INSA Senior scientist position at NISER, Bhubaneswar through the Grant number INSA/SP/SS/2023. Last but not least, we acknowledge our debt to the people
belonging to the various parts of the world for their generous and steady support for research in natural sciences.
\newpage
\section*{Appendix}
\appendix
\section{ Bogoliubov coefficients}\label{bogoCoeff}
The Bogoliubov coefficients which describe the USR and SRII phases are given by the following expression:
\begin{widetext}
    \bea \label{alpha2b}
\alpha_{{2}} &=&  \frac{1}{2 k^3 \tau_s^3 c_s^3}  \Bigg(3 i + 3 i  k^2 c_s^2 \tau_s ^2  + 2  k^3 c_s^3  \tau_s ^3  \Bigg  ),\\ \label{beta2b}  \beta_{{2}} &=& \frac{1}{2 k ^3 c_s ^3 \tau_s ^3}  \Bigg( 3i -6 k c_s \tau_s -3i k^2 c_s ^2 \tau_s^2 \Bigg) e^{-i\left(\pi\left(\nu+\frac{1}{2}\right)+ 2k c_s \tau_s\right)},
\\
 \alpha _{3} &=& \frac{1}{(2 k^3 \tau_e^3 c_s^3)(2 k^3 \tau_s^3 c_s^3)}\Bigg\{\left(-3 i -3 i  k^2 \tau_e^2 c_s^2 +2  k^3 \tau_e^3 c_s^3 \right)\times\left(3 i + 3 i  k^2 c_s^2 \tau_s ^2  + 2  k^3 c_s^3  \tau_s ^3  \right  ) \nonumber\\ && \quad\quad\quad\quad- \left(-3 i -6  k \tau_e c_s   +3 i  k^2 \tau_e^2 c_s^2 \right)\times
 \left( 3i -6 k c_s \tau_s -3i k^2 c_s ^2 \tau_s^2 \right) e^{2 i k c_s( \tau_e -\tau_s) }\Bigg\},
\\
  \beta _{3} &=&  \frac{1}{(2 k^3 \tau_e^3 c_s^3)(2 k ^3 c_s ^3 \tau_s ^3)}\Bigg\{ \left(-3 i  +6  k \tau_e c_s +3 i k^2 \tau_e^2 c_s^2\right) \times\left(3 i + 3 i  k^2 c_s^2 \tau_s ^2  + 2  k^3 c_s^3  \tau_s ^3  \right) e^{-\left(2 i k \tau_e c_s + i \pi  \left(\nu +\frac{1}{2}\right)\right)} \nn\\&&\quad\quad\quad\quad +\left(2  k^3 \tau_e^3 c_s^3 + 3 i  k^2 \tau_e^2 c_s^2 +3 i \right )  \times\left( 3i -6 k c_s \tau_s -3i k^2 c_s ^2 \tau_s^2 \right) e^{-i\left(\pi(\nu+\frac{1}{2})+ 2k c_s \tau_s \right)}\Bigg\}.
\eea 
\end{widetext}

\section{Special functions}
\label{A1}
The transfer function in the expression for the SIGW spectrum is expressed in terms of associated Legendre functions $P_\nu^\mu(x)$  and $Q_\nu^\mu(x)$ which are given by:
\bea
P_\nu^\mu(x)&=&\left[\frac{1+x}{1-x}\right]^{\mu/2}{\bf F}(v+1,-v;1-\mu;\frac{1}{2}(1-x)),\quad\quad\\
Q_{\nu}^{\mu}(x) &=& \frac{\pi}{2\sin{(\mu\pi)}} \bigg[\cos{(\mu \pi)}\bigg(\frac{1+x}{1-x}\bigg)^{\mu /2} \nonumber\\&&{\bf F}\bigg(v+1,-v;1-\mu;\frac{1}{2}(1-x)\bigg)\nonumber\\
  &&- \frac{\Gamma (v+\mu +1)}{\Gamma (v-\mu +1)}\bigg(\frac{1-x}{1+x}\bigg)^{\mu /2} \nonumber\\&&{\bf F}\bigg(v+1,-v;1+\mu;\frac{1}{2}(1-x)\bigg )\bigg]
.
\eea
Here ${\bf F}(a,b;c;x) = F(a,b;c;x)/\Gamma(c)$ where $F(a,b;c;x)$ is the Gauss's Hypergeometric Function that is scaled with $\Gamma(c)$ to obtain the new function ${\bf F}(a,b;c;x)$. These above functions are known as Ferrer's functions and their validity extends to $|x|<1$. However when $|x|>1$, we have Oliver's functions given by:
\bea
\label{calQ}
Q_{\nu}^{\mu}(x) &=& \frac{\pi}{2 \sin{\mu \pi}\Gamma(\nu + \mu +1)}\bigg[\bigg(\frac{x+1}{x-1}\bigg)^{\mu /2}\nonumber\\&&{\bf F}\bigg(v+1,-v;1-\mu;\frac{1}{2}-\frac{1}{2}x \bigg)\nonumber \\
&&   -\frac{\Gamma (v+\mu +1)}{\Gamma (v-\mu +1)}\bigg(\frac{x-1}{x+1}\bigg)^{\mu /2}\nonumber\\&&{\bf F}\bigg(v+1,-v;1-\mu;\frac{1}{2}(1-x)\bigg) \bigg].
\eea

\bibliography{RefsGWB}
\bibliographystyle{utphys}
\end{document}